\documentclass[preprint,aps,prc,showpacs]{revtex4}

\usepackage{graphicx}
\usepackage{bm}

\begin{document}

\title{ Dynamical simulation of bound antiproton-nuclear systems and 
        observable signals of cold nuclear compression }
\author{A.B. Larionov$^{1,2}$, I.N. Mishustin$^{1,2}$, L.M. Satarov$^{1,2}$,
        and W. Greiner$^1$}

\affiliation{$^1$Frankfurt Institute for Advanced Studies, J.W. Goethe-Universit\"at,
             D-60438 Frankfurt am Main, Germany\\ 
             $^2$Russian Research Center Kurchatov Institute, 
             123182 Moscow, Russia}

\date{\today}

\begin{abstract}
On the basis of the kinetic equation with selfconsistent relativistic 
mean fields acting on baryons and antibaryons, we study dynamical 
response of the nucleus to an antiproton implanted in its interior.
By solving numerically the time-dependent Vlasov equation, we show that 
the compressed state is formed on a rather short time scale of about 
$4\div10$ fm/c. This justifies the assumption, that the antiproton 
annihilation may happen in the compressed nuclear environment.
The evolution of the nucleus after antiproton annihilation is described
by the same kinetic equation including collision terms. 
We show, that nucleon kinetic energy spectra and the total invariant mass 
distributions of produced mesons are quite sensitive observables to the 
antiproton annihilation in the compressed nucleus.
\end{abstract}


\maketitle

\section{ Introduction }

As has been shown recently in Refs. \cite{Buer02,Mish05}, an antiproton
implanted in a heavy nucleus serves as an attractor for surrounding 
nucleons that can lead to a sizable increase of the central nucleon density. 
This effect is caused by the strong attractive scalar and vector potentials 
acting on the antiproton, as follows from the $G$-parity transformation of 
nuclear potentials \cite{GM01}. Correspondingly, the antiproton also creates 
an attractive potential acting on nucleons. This leads to the concentration 
of nucleons around the antiproton and, as result, to a considerable increase
of the nucleon density.

Within the relativistic mean field (RMF) model, the $G$-parity transformed 
nuclear optical potential is about $-700$ MeV at the normal nuclear matter
density $\rho_0=0.148$ fm$^{-3}$, while a phenomenological value of an antiproton 
optical potential is limited within the range of $-(100\div350)$ MeV 
\cite{Wong84,Teis94,Batty97,Friedman05,Friedman07}. Therefore, in order to fit 
the empirical optical potential, the antiproton coupling constants with
$\sigma$-, $\omega$- and $\rho$-meson fields should be reduced 
with respect to the values given by the $G$-parity transformation.
The RMF calculations with reduced coupling constants \cite{Mish05} 
still show quite strong compressional effects for light and medium nuclei.  

An important question, which arises here is whether the compression process
is fast enough to develop before the $\bar p$-annihilation. The total $\bar p p$%
-annihilation cross section in vacuum can be parameterized at low relative velocities 
$v_{\rm rel}$ as
\begin{equation}
   \sigma^{\bar p p}_{\rm ann} = C + \frac{D}{v_{\rm rel}}~,     \label{sigma_ann}
\end{equation}
where $C=38$ mb and $D=35$ mb$\cdot$c \cite{Dover92}. Using these 
numbers we can estimate the life time of an antiproton inside the
nuclear matter at normal density:
\begin{equation}
   \tau_{\rm ann} \simeq \frac{1}{ \rho_0\, \sigma^{\bar p p}_{\rm ann}\, 
                                    v_{\rm rel}} \simeq 2~\mbox{fm/c}~.
                                                       \label{tau_ann}
\end{equation}
This is, of course, a very short time in nuclear scale. 
However, as argued in Ref. \cite{Mish05}, this time can become much longer,
up to 20 fm/c, for deeply bound antiprotons due to the phase space suppression
factors. Therefore, the compression effects can, in-principle, show up in
$\bar p$-nuclear interactions.

In the present work, we apply a dynamical transport model in order to study
the formation and decay of the compressed $\bar p$-nuclear system. Our 
calculations are based on the Giessen Boltzmann-Uehling-Uhlenbeck (GiBUU) model 
\cite{GiBUU}, which has been recently supplemented by the relativistic mean fields 
\cite{LBGM07}. Apart from collision terms, the GiBUU model solves the coupled 
(through the mean fields) Vlasov equations for nucleon and antiproton phase space 
distribution functions.
As well known \cite{Wong82}, the Vlasov equation provides a semiclassical limit 
of the time-dependent Hartree-Fock calculations. Thus, the compressional effects 
found in Refs. \cite{Buer02,Mish05} should also be reproduced as a static solution 
of the coupled Vlasov equations. 

It will be demonstrated that the compression process is characterized by the time 
scale which is comparable with the $\bar p$ life time in nuclear environment.
Thus, $\bar p$ has, indeed, a chance to annihilate inside the compressed nucleus.
We will show, that the $\bar p$-annihilation in a compressed nucleus should lead
to the collective expansion of the residual nuclear system. The appearance of
the high-energy tails in the kinetic energy spectra of the emitted nucleons
is predicted. The distributions in the total invariant mass of 
produced mesons reveal a noticeable shift toward lower invariant masses, when the 
annihilation takes place inside the compressed nucleus.

The annihilation of slow antiprotons inside heavy nuclei was, first, proposed
by Rafelski \cite{Rafelski80} as a unique opportunity to study nuclear matter in 
unusual conditions.
Later, Cahay et al. \cite{Cahay82} studied the $\bar p$ annihilation inside nuclei
within an intranuclear cascade model.
In Ref. \cite{Cahay82}, antiproton annihilation events into pions at the center
of $^{40}$Ca and $^{108}$Ag nuclei were simulated. The mean field effects
were, however, completely neglected in \cite{Cahay82}. 

In Sect. II, we describe the theoretical model applied in calculations.
Sect. III contains the results of the time evolution study for the compression
and explosion dynamics. In Sect. IV, we propose several observable signals
sensitive to the $\bar p$-annihilation in the compressed nucleus.
The summary and outlook are given in Sect. V.  

\section{The model}

In calculations, we apply the GiBUU model developed in Giessen University.
For the detailed description and related references, we refer the reader to the 
web page \cite{GiBUU}, where the new version of the model is presented. 
Below, we mostly describe the new features implemented in the present work.

\subsection{Relativistic mean fields} 

Below we consider a system composed of an antinucleon interacting with baryons. 
This system is described by the RMF Lagrangian of the following form \cite{Mish05,LKR97}:
\begin{eqnarray}
      {\cal L} & = & \sum_{j=B,\bar N} \bar\psi_j [ \gamma_\mu ( i\partial^\mu - g_{\omega j}\omega^\mu ) 
      - m_j - g_{\sigma j} \sigma ] \psi_j                                           \nonumber \\
               & + & {1 \over 2}\partial_\mu\sigma\partial^\mu\sigma - U(\sigma)
                 -  {1 \over 4} F_{\mu\nu} F^{\mu\nu} + {1 \over 2} m_\omega^2 \omega_\mu\omega^\mu~,    
                                                                                         \label{Lagr}
\end{eqnarray}
where $\psi_j$ are the baryon ($j=B \equiv N,~N^\star,~\Delta,~Y$) and antinucleon 
($j=\bar N$) fields, respectively; 
$\sigma$ is the isoscalar-scalar meson field ($I^G=0^+$, $J^\pi=0^+$);
$\omega^\mu$ is the isoscalar-vector meson field ($I^G=0^-$, $J^\pi=1^-$);
and $F_{\mu\nu} \equiv \partial_\mu \omega_\nu - \partial_\nu \omega_\mu$.
Here $N^\star$ and $\Delta$ denotes, respectively, the isospin 1/2 and 3/2 nonstrange
baryonic resonances, and $Y$ stands for the $S=-1$ baryons explicitly propagated in
the GiBUU model \cite{GiBUU}. In the case of the spin 3/2, 5/2 and 7/2 baryonic 
resonances, their fields $\psi_j$ carry also one or more vector
indices, which are dropped in Eq.(\ref{Lagr}) and below for brevity. 
When appropriate, the covariant summation is assumed over these indices.
For simplicity, the isovector and electromagnetic terms are disregarded in (\ref{Lagr}).
The selfinteractions of the $\sigma$-field are included in (\ref{Lagr})
via the term $U(\sigma)$ in order to avoid an unrealistically high compressibility
coefficient of the nuclear matter \cite{Boguta77}:
\begin{equation}
      U(\sigma) = {1 \over 2} m_\sigma^2 \sigma^2 +  {1 \over 3} g_2 \sigma^3 
                                                  +  {1 \over 4} g_3 \sigma^4~.  \label{U}
\end{equation}

Some comments are in order to gain more insight into Eq. (\ref{Lagr}).
Following Ref. \cite{Mish05}, the antinucleon field $\psi_{\bar N}$ in the Lagrangian 
density (\ref{Lagr}) is represented in terms of wave functions of
{\it physical antinucleons}. These wave functions can be obtained by the 
$G$-parity transformation acting on the wave functions of the Dirac sea nucleons 
(see Ref. \cite{GM01} for details), which appear in the relativistic description
of the nucleon \cite{Walecka74}. By applying the same transformation, the nonlinear RMF
Lagrangian of Refs. \cite{Boguta77,LKR97} (neglecting terms responsible for 
the baryon-antibaryon annihilation) can be expressed as (\ref{Lagr}) with the following 
relations between coupling constants:
\begin{equation}
   g_{\omega \bar N} = -g_{\omega N},~~~~~g_{\sigma \bar N} = g_{\sigma N}~.
                                                                   \label{Gpar_relation}
\end{equation}
The relations (\ref{Gpar_relation}) are satisfied if the physical system would be 
exactly symmetric with respect to the $G$-parity transformation. However, this
is not necessary to be true in a many-body system \cite{Mish05,GM01}. 
The reason is that the concept of the $G$-parity symmetry is strictly applicable
on the level of the elementary processes only. However, the RMF Lagrangian 
(\ref{Lagr}) is dealing with the effective interactions, which are usually tuned 
to describe the bulk properties of the nuclear medium and/or the properties of some 
selected nuclei. Due to the many-body effects, such as the Pauli blocking or
mixed scalar-vector terms in the scattering amplitudes, these effective interactions
may not obey the exact $G$-parity symmetry anymore. To take into account possible 
deviations from the $G$-parity symmetry, we introduce an overall scaling of the 
antinucleon-meson coupling constants with respect to the values given by 
(\ref{Gpar_relation}) (see Ref. \cite{Mish05}):
\begin{equation}
   g_{\omega \bar N} = -\xi g_{\omega N},
   ~~~~~g_{\sigma \bar N} = \xi g_{\sigma N}~,            \label{scaled_relation}
\end{equation}
where $0 < \xi \leq 1$ is a scaling factor.

Throughout the paper, we consider two options for the scaling factor
of the antinucleon-meson coupling constants:
$\xi = 1$, motivated by the $G$-parity, and $\xi=0.3$, which is in a better 
agreement with the empirical $\bar p$A optical potential.
For other baryonic fields we put 
in the present work, for simplicity, the same coupling constants as for the nucleon~:
$g_{\omega N^*} = g_{\omega \Delta}  = g_{\omega Y} = g_{\omega N}$,
$g_{\sigma N^*} = g_{\sigma \Delta} = g_{\sigma Y} = g_{\sigma N}$.

All calculations have been performed emloying the NL3 parameterization \cite{LKR97}
of the RMF model. This parameterization provides quite reasonable nuclear matter
properties: the binding energy 16.299 MeV/nucleon, the compressibility 
coefficient $K=271.76$ MeV and the nucleon effective mass $m_N^*=0.60\, m_N$ 
at $\rho_0$. Moreover, the NL3 parameterization reproduces the ground state properties 
of spherical and deformed nuclei very well \cite{LKR97}.

The Dirac equations of motion for baryons have the following form:
\begin{equation}
   ( \gamma^\mu ( i\partial_\mu - g_{\omega j} \omega_\mu ) - m_j^\star ) \psi_j = 0~,
                                                                    \label{DiracEqB} 
\end{equation}
where 
\begin{equation}
   m_j^\star = m_j + g_{\sigma j} \sigma           \label{mstar}
\end{equation}
is the effective (Dirac) mass.

Within the mean field approximation the $\sigma$- and $\omega$-fields are treated classically.
They satisfy the (nonlinear) Klein-Gordon-like equations with the source terms due to 
coupling to baryons and an antinucleon:
\begin{eqnarray}
     & & \partial_\nu\partial^\nu \sigma + { \partial U(\sigma) \over \partial \sigma } 
                               = - \sum_{j=B,\bar N} g_{\sigma j} \rho_{S j}~,   \label{KGeqSigma} \\
     & & (\partial_\nu\partial^\nu + m_\omega^2)\,\omega^\mu 
                               = \sum_{j=B,\bar N} g_{\omega j} j_{b j}^\mu~,    \label{KGeqOmega}
\end{eqnarray}
where $\rho_{S j} = < \bar\psi_j \psi_j >$ is the partial scalar density 
and $j_{b j}^\mu = < \bar\psi_j \gamma^\mu  \psi_j >$ is the partial baryon current. 
Equation (\ref{KGeqOmega}) has to be supplemented
by the four-transversality condition 
\begin{equation}
   \partial_\mu \omega^\mu=0~.                                              \label{fourTrans}
\end{equation}

\subsection{ Covariant kinetic equations }

Instead of solving the Dirac equations (\ref{DiracEqB}), we will describe 
the baryons and antinucleon dynamics by the coupled set of the semiclassical 
kinetic equations \cite{Vasak87,Elze87,Ivanov87,Blaettel93,LBGM07}:
\begin{equation}
  \frac{1}{p_0^\star}
  \left[ p^{\star\mu} \frac{\partial}{\partial x^\mu} + \left(g_{\omega j} p_\mu^\star F^{k\mu} 
                                   + m_j^\star \frac{\partial m_j^\star}{\partial x_k}\right)
    \frac{\partial}{\partial p^{\star k}} \right] f_j(x,{\bf p^\star}) 
  = I_j[f_B,f_M]~,       
                                                                                \label{kinEq}
\end{equation}
where $k=1,2,3$; $\mu=0,1,2,3$; $x \equiv (t,{\bf r})$; and $f_j(x,{\bf p^\star})$  is the 
distribution function (DF) in a six-dimensional phase space $({\bf r, p^\star})$ with 
${\bf p^\star}$ being the spatial components of the kinetic four-momentum
\begin{equation}
   p^{\star \mu} = p^\mu - g_{\omega j} \omega^\mu~.                           \label{pKin}
\end{equation}
The baryons and antinucleon are assumed to be on the respective effective mass shells:
\begin{equation}
   p^{*0} = \sqrt{ ({\bf p^\star})^2 + (m_j^\star)^2 }~.                        \label{massShell}
\end{equation}
The l.h.s. of Eq. (\ref{kinEq}) describes the propagation of the $j$-th type particles in the
classical $\sigma$- and $\omega$-fields. The r.h.s. of Eq. (\ref{kinEq}) is a collision integral,
which represents the (in)elastic two-body collisions with corresponding vacuum cross sections 
as well as the resonance decays.
The complete description of the collision integral
structure, in-particular, the differential elementary cross sections included
into the GiBUU model can be found in \cite{GiBUU,LBGM07} and in refs. therein.
The in-medium modification of the baryon-baryon and baryon-meson 
cross sections is neglected in the present work. 

We will apply the full kinetic equations, including collision terms, only to describe
the {\it post-annihilation} evolution of a system.  By this reason, the antiproton DF is 
excluded from the collision integral. Instead, we enforce $\bar p$ to annihilate into 
mesons at some preselected time (see Sect. IIE).    
Thus, in the present work the collision integral includes the nucleon, $\Delta(1232)$ and 
higher baryon resonances up to the mass of 2 GeV, which can be excited in the meson-baryon
and baryon-baryon collisions. 
A possible hyperon formation in the processes $\pi N \to Y K$ and $\bar K N \to \pi Y$
is included too. The ``valence mesons'' $M \equiv \pi,~\eta,~\rho,~\sigma,~\omega,~\eta^\prime,~
\phi,~\eta_c,~J/\psi,~K,~\bar K,~K^*,~\bar K^*$ are explicitly taken into account.
They are assumed to propagate freely between collisions, i.e. we neglect the mean field potentials 
acting on these mesons. 

The scalar density and the baryon current of the $j$-th type baryons are expressed in terms of 
DF as follows:
\begin{eqnarray}
\rho_{S j}(x) & = & \frac{g_j}{(2\pi)^3} \int\,\frac{ d^3 p^\star }{ p^{\star0} } m_j^\star
               f_j(x,{\bf p^\star})~,                                         \label{rhoSj} \\
j_{b j}^\mu(x) & = & \frac{g_j}{(2\pi)^3} \int\,\frac{ d^3 p^\star }{ p^{\star0} } p^{\star\mu}
               f_j(x,{\bf p^\star})~,                                         \label{jBj}
\end{eqnarray}
where $g_j$ is the spin-isospin degeneracy factor ($g_N = g_{\bar N} = 4~,~g_\Delta=16$ etc.).

One can show \cite{Blaettel93,Weber92} that the kinetic equations (\ref{kinEq}) with the
$\sigma$- and $\omega$-fields evolving according to Eqs. (\ref{KGeqSigma}),(\ref{KGeqOmega})
lead to the continuity equations
\begin{equation}
   \sum_{j=B}\partial_\mu j_{b j}^\mu = 0~,~~~~ \partial_\mu j_{b \bar N}^\mu = 0   \label{contEq}
\end{equation}
and the energy-momentum conservation
\begin{equation}
   \partial_\nu T^{\mu \nu} = 0~,                                \label{enmom}   
\end{equation}
where the energy-momentum tensor is written as
\begin{eqnarray}
  & &    T^{\mu \nu} =   
         \sum_{j=B,\bar N,M} 
         \frac{g_j}{(2\pi)^3} \int\,\frac{ d^3 p^\star }{ p^{\star0} }
         p^\mu p^{\star \nu} f_j(x,{\bf p^\star})                 
       + \partial^\mu\sigma \partial^\nu\sigma
       - \partial^\mu\omega^\lambda\partial^\nu\omega_\lambda            \nonumber \\
  & &  - g^{\mu \nu} \left( \frac{1}{2}\partial_\lambda\sigma\partial^\lambda\sigma  - U(\sigma)
                          - \frac{1}{2} \partial_\lambda\omega_\kappa\partial^\lambda\omega^\kappa
                          + \frac{1}{2} m_\omega^2 \omega^2 \right)~.     \label{T}
\end{eqnarray}
Here we have also included possible contributions of the ``valence'' mesons $M$, which
can be produced at the annihilation. It is assumed that $p^\star = p$ for the valence mesons.

\subsection{Numerical realization}

In order to solve Eq. (\ref{kinEq}) numerically, DF is represented by the set of point-like
test particles:
\begin{equation}
   f_j(x,{\bf p}^\star) = \frac{ (2\pi)^3 }{ g_j n } \sum_{i=1}^{n N_j} \delta( {\bf r} - {\bf r}_i(t) )
                                                         \delta( {\bf p}^\star - {\bf p}^\star_i(t) )~, 
                                                                                \label{tP}
\end{equation}
where $N_j$ is the number of physical particles of the type $j$ and $n$ is the number of
test particles per physical particle (the same for all types $j$). The test particle
positions ${\bf r}_i$ and kinetic momenta ${\bf p}^\star_i$ are evolving in time 
according to the following equations: 
\begin{eqnarray}
& & \dot{\bf r}_i =  { {\bf p}_i^\star \over p_i^{\star 0} }~,                      \label{rDot} \\
& & \dot{p}_i^{\star k} = g_{\omega j} { p_{i\mu}^\star \over p_i^{\star 0} } F^{k\mu}
                              + { m_j^\star \over p_i^{\star 0} } 
                                \frac{\partial m_j^\star}{\partial x_k}             \label{pStarDot}
\end{eqnarray}
with $k=1,2,3$ and $\mu=0,1,2,3$. It is easy to check that DF (\ref{tP}) with ${\bf r}_i$ and
${\bf p}_i^\star$ satisfying Eqs. (\ref{rDot}),(\ref{pStarDot}) gives a formal solution
of the Vlasov equation in the case when the collision integral in (\ref{kinEq}) is equal 
to zero. Equations (\ref{rDot}),(\ref{pStarDot}) are equivalent to the Hamiltonian 
equations of motion for the test particle positions ${\bf r}_i$ and 
canonical momenta ${\bf p}_i$:
\begin{eqnarray}
& & \dot{\bf r}_i =  \frac{\partial p_i^0}{\partial {\bf p}_i}~,       \label{rDotHam} \\
& & \dot{\bf p}_i = -\frac{\partial p_i^0}{\partial {\bf r}_i}~,       \label{pDotHam}
\end{eqnarray}
where $p_i^0 = g_{\omega j} \omega^0 + \sqrt{ ({\bf p}_i^\star)^2 + (m_j^\star)^2 }$ is
the single-particle energy (see \cite{Blaettel93,Ko87}). However, it is more
convenient to propagate in time the test particle kinetic momenta rather than the canonical 
ones, since then Eq.~(\ref{KGeqSigma}) for the $\sigma$-field decouples from
Eq.~(\ref{KGeqOmega}) for the $\omega$-field.

When the collision integral in Eq. (\ref{kinEq}) is taken into account, the test particles
are propagated between the two-body collisions using Eqs. (\ref{rDot}),(\ref{pStarDot}).
All calculations have been performed in the parallel ensemble mode. In this mode, the two-body 
collisions are permitted between the test particles belonging to the same parallel ensemble only, 
while the mean field is averaged over $n$ parallel ensembles of the test particles propagated 
simultaneously (see Eq. (\ref{tP})). Therefore, a single parallel ensemble can be considered 
as a physical event.  

In actual calculations, we have neglected the time derivatives of the meson fields in 
Eqs. (\ref{KGeqSigma}),(\ref{KGeqOmega}). However, the spatial derivatives were treated
without any simplifying assumptions. The reason for such a strategy is that we are dealing
with nuclear systems which have large density gradients, but evolving slowly, as compared
with the spatial and temporal scales involved in the mesonic equations of motion. Indeed, including the temporal gradients would lead to the frequent 
oscillations of the mesonic fields with a period of less than 
$2\pi/m$, where $m$ is the meson mass. This gives the period of 
2.5 fm/c (1.5 fm/c) for the $\sigma$- ($\omega$-) field. By taking into 
account the finite wave lenghts of these oscillations would further
reduce the periods. The treatment of such oscillations would strongly 
complicate the numerical calculations, in-particular, due to the classical 
meson field radiation. On the other hand, the characteristic periods of the 
oscillations are significantly smaller than the characteristic
compression times ($4\div10$ fm/c, see Sect. III A below). Therefore, one can
approximately average-out the mesonic fields with respect to these 
oscillations, that is actually assumed in our model. 
The $\sigma$- and $\omega$-fields are, therefore, calculated from the equations
\begin{eqnarray}
     & & -\triangle\sigma + { \partial U(\sigma) \over \partial \sigma }
                               = - \sum_{j=B,\bar N} g_{\sigma j} \rho_{S j}~, \label{eqSigma} \\
     & & (-\triangle + m_\omega^2)\,\omega^\mu 
                               =  \sum_{j=B,\bar N} g_{\omega j} j_{b j}^\mu~.  \label{eqOmega}
\end{eqnarray}
Within the same approximation, the energy-momentum tensor has the following form:
\begin{eqnarray}
  & &    T^{\mu \nu} = \sum_{j=B,\bar N,M} 
          \frac{g_j}{(2\pi)^3} \int\,\frac{ d^3 p^\star }{ p^{\star0} }
         p^\mu p^{\star \nu} f_j(x,{\bf p^\star})                  
       + (  \partial^\mu\sigma \partial^\nu\sigma
          - \partial^\mu\omega^\lambda\partial^\nu\omega_\lambda )
         (1-\delta_{\nu 0})                                        \nonumber \\
  & &  - g^{\mu \nu} \left( -\frac{1}{2}(\nabla\sigma)^2 - U(\sigma)
                            +\frac{1}{2} \nabla\omega_\lambda\nabla\omega^\lambda
                            +\frac{1}{2} m_\omega^2 \omega^2 \right)~.     \label{Twotd}
\end{eqnarray}
The factor $(1-\delta_{\nu 0})$ in Eq. (\ref{Twotd}) reflects the fact, that
due to the omission of the time derivatives of the meson fields in the Lagrangian density,
only the first term in the r.h.s. contributes to the three-momentum density 
$T^{\alpha 0}$  ($\alpha=1,2,3$).

Although Eqs. (\ref{eqSigma}),(\ref{eqOmega}) are not covariant, they provide 
a better description of the nuclear surface than pure local fields \cite{LBGM07}.
This improves the stability of a nuclear ground state and is more appropriate
for studying nuclear response to external hadronic and electromagnetic probes.

Equations (\ref{eqSigma}),(\ref{eqOmega}) have been solved numerically by applying the
alternating direction implicit iterative method of Douglas described in  Ref. 
\cite{Varga62}. Due to the scalar density dependence on the effective mass 
(see Eqs.(\ref{mstar}),(\ref{rhoSj})), additional iterations are needed to solve 
Eq.(\ref{eqSigma}). In other words, the scalar density has to be computed selfconsistently.  
To evaluate the meson fields, we used a uniform grid in coordinate 
space with steps $\Delta_x=\Delta_y=\Delta_z$. For the systems $\bar p$$^{16}$O, 
$\bar p$$^{40}$Ca and $\bar p$$^{208}$Pb considered below, the grid covered a cubic
volume with the side of 10, 20 and 30 fm, respectively, centered at the center-of-mass 
(c.m.) of a $\bar p$A system.
By numerical reasons, the $\delta$-functions in coordinate space, 
introduced in Eq. (\ref{tP}), have been replaced by the Gaussians of the width $L$:
\begin{equation}
   \delta( {\bf r} - {\bf r}_i(t) ) \Rightarrow \frac{1}{(2\pi)^{3/2}L^3}
   \exp\left\{-\frac{({\bf r} - {\bf r}_i(t))^2}{2L^2}\right\}~.       \label{gauss}
\end{equation}
The width of the Gaussian and the grid step sizes are pure numerical parameters which
should resolve the coordinate space nonuniformities of the system. In our
case the characteristic space scale is given by the radius of the smallest considered
nucleus $^{16}$O, i.e. $\simeq3$ fm. On the other hand, in order to have smooth
density distributions, the number $n$ of test particles per physical particle 
(see Eq. (\ref{tP})) should be correlated to the width of the Gaussian as
$n \propto L^{-3}$. This puts a restriction on too small width due to CPU
time increase. As an optimum choice, we fixed in the present work     
$\Delta_x=\Delta_y=\Delta_z=L=0.5$ fm.  We have realised, however, that there 
is a rather moderate tendency of increasing maximum compression (see discussion
in Sect. III A below) with decreasing Gaussian width. The number of test particles 
per nucleon was set to $n=1500$ in the most of calculations.

The equations of motion (\ref{rDot}),(\ref{pStarDot}) have been solved by applying the second-order 
in time predictor-corrector method \cite{LBGM07} with the time step of 0.1 fm/c. This value
is small enough to resolve the time scale of a few fm/c for the compression processes
(see Figs. \ref{fig:rho_U_vs_z} and \ref{fig:rho_vs_t} below). We have checked, that taking smaller
time step does not influence the results. The full numerical scheme conserves the total energy with the
accuracy of about 5\% of the initial total binding energy of the $\bar p$A system.

\subsection{ Initialization }

The nucleons were distributed in coordinate space according to the Woods-Saxon
density profile. The momenta of nucleons were sampled according to the local Fermi
distribution. 

The initial antiproton DF was chosen as a Gaussian wave packet in
coordinate and momentum space \cite{Aichelin91,Bondorf97} located at the center of 
a nucleus ($x=y=z=0$):
\begin{equation}
   f_{\bar N}(t=0,{\bf r},{\bf p^\star}) = \frac{ (2\pi)^3 }{ g_{\bar N} \pi^3 }
   \exp\{ -{\bf r}^2/(2\sigma_r^2) - 2\sigma_r^2{\bf p^\star}^2 \}~,    \label{cohState}
\end{equation}
where $\sigma_r$ is the width in coordinate space. Equation (\ref{cohState}) implies, that 
the antiproton is at rest. The width of the initial antiproton distribution 
in momentum space is $(2\sigma_r)^{-1}$, which follows from the uncertainty relation. 
If not mentioned explicitly, the calculation is
done with the choice $\sigma_r=1$ fm. This value agrees with results of the 
static RMF calculations of Ref. \cite{Mish05}.  
As for nucleons, the antiproton DF (\ref{cohState})
is projected onto test particles according to Eq. (\ref{tP}) with the
$\delta$-functions in coordinate space replaced by Gaussians.
To avoid misunderstanding, we note that one should distinguish the width $\sigma_r$ 
of the physical antiproton spatial distribution in Eq. (\ref{cohState}) and 
the width of the test particle Gaussian.

\subsection{ Propagation and annihilation }

After the initialization, the system of nucleons and antiproton was propagated
in time according to Eqs. (\ref{rDot}),(\ref{pStarDot}). The meson fields
have been calculated by Eqs. (\ref{eqSigma}),(\ref{eqOmega})
with the source terms given by the scalar densities (\ref{rhoSj}) and the baryon currents
(\ref{jBj}). In such a way, the evolution of the system toward compressed state
has been followed. In this calculation, the collision term in the r.h.s. of the 
kinetic equation (\ref{kinEq}) has been set to zero, i.e. we considered
a pure mean-field Vlasov dynamics. This was done to see most clearly the role
of the mean fields. An introduction of the $\bar N N$ and $NN$ elastic collisions would 
mainly lead to a dissipation of the collective energy into heat.
As pointed out in Ref. \cite{Mish05}, this effect is rather small and, therefore, can not 
change significantly the compression dynamics.  

The reason is, that the elastic collisions are not frequent on the time scale
of compression (see Figs. \ref{fig:rho_U_vs_z}, \ref{fig:rho_vs_z} and \ref{fig:rho_vs_t}).
Indeed, the mean time $\tau_{\rm coll}$ between nucleon-nucleon collisions  can be estimated 
as $\tau_{\rm coll}=1/(\rho_N \sigma_{NN} v_F)$, where $\sigma_{NN} \simeq 40$ mb is
the elastic nucleon-nucleon cross section (c.f. Refs. \cite{Li93,Cugnon96}) 
and $v_F \simeq 0.3c$ is the Fermi velocity. This gives $\tau_{\rm coll}=3\div6$ fm/c 
for the nucleon density $\rho_N=2\div1 \rho_0$.
The Pauli blocking effect will further increase $\tau_{\rm coll}$. A similar
estimate can also be done for $\bar N N$ elastic collisions.

At certain time moment $t_{\rm ann}$, which is an external parameter to our model, 
we simulated the annihilation of an antiproton. This implies that annihilation 
occurs instantaneously, as a single quantum mechanical transition, in 
distinction to description of this process via the collision term in a kinetic 
equation. In the last case, the antiproton distribution function would gradually
disappear on the way to the compressed state. The purpose of the present work
is to look at the strongest possible effect of the nuclear compression on 
observables. Therefore, we let the compressed system to be formed, and simulate
the sudden annihilation afterwards. The ambiguity in the in-medium annihilation cross
sections is taken into account by varying the parameter $t_{\rm ann}$.  

In the actual calculations, the annihilation was simulated as follows: For each antiproton test 
particle, the closest in coordinate space nucleon test particle was chosen to be the 
annihilation partner. At large enough values of the total in-medium c.m. energy $\sqrt{s}$
of the annihilating $\bar p N$ pair (see below), the annihilation event of the test particle 
pair into mesons was simulated using the quark model \cite{GCM98,Geiss98}, which has been
already implemented in the GiBUU model \cite{GiBUU} earlier. A quark and an antiquark 
with the same flavour are assumed to annihilate and transfer their total four-momentum 
to the remaining (anti)quarks. The remaining four (anti)quarks form two orthogonal $q \bar q$ jets 
with equal energies in the c.m. frame. The jets were hadronized via the Lund string fragmentation 
model \cite{Lund} in the JETSET version included into the PYTHIA 6.225 program package.
The applied annihilation model corresponds to the R2 type diagram in classification of 
Ref. \cite{Dover92}, i.e. to the quark rearrangement with one $q\bar q$ annihilation vertex. 
In this sense, the model has some similarity with the two-meson doorway models of Refs. 
\cite{Vandermeulen88,Mundingl91}. To illustrate how the model works we have performed simulations 
of the $p \bar p$ annihilation.

\begin{figure}

\includegraphics[bb = 106 150 589 503]{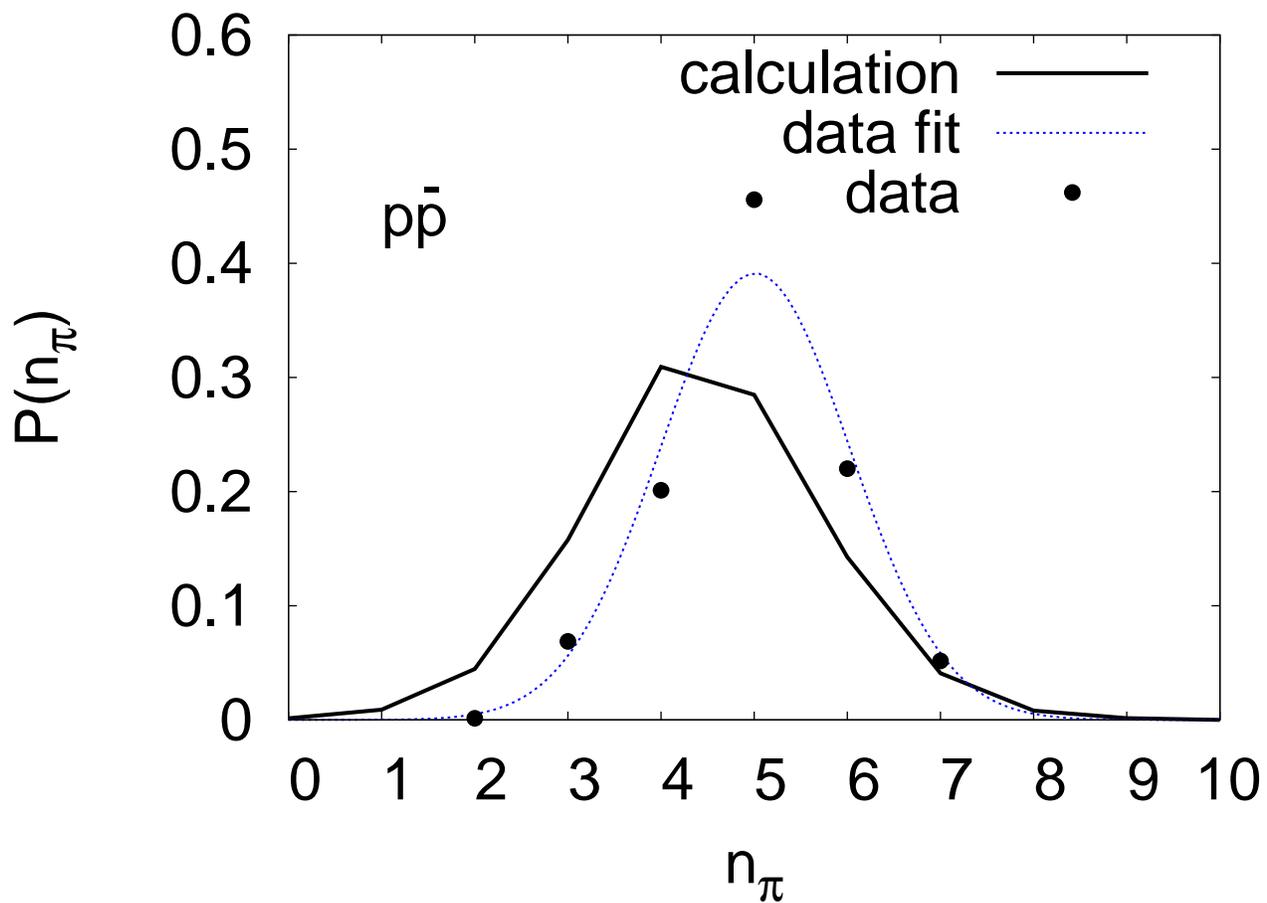}

\vspace*{1cm}

\caption{\label{fig:pimul_ppbar} (color online) Pion multiplicity distribution
for $p \bar p$ annihilation at rest in vacuum. Data points are from 
Ref. \cite{Dover92}. The dashed line represents the data fit \cite{SS88} with the 
Gaussian $P(n_\pi)=\exp\{-(n_\pi-<n_\pi>)^2/2\sigma_{n_\pi}^2\}/\sqrt{2\pi\sigma_{n_\pi}^2}$
where $<n_\pi>=5.01$ and $\sigma_{n_\pi}^2=1.04$.}
 
\end{figure}

Fig.~\ref{fig:pimul_ppbar} shows the pion multiplicity distribution for
the $p \bar p$ annihilation at rest in vacuum compared to the data compilation 
from Refs. \cite{Dover92,SS88}. The calculated distribution is somewhat shifted to smaller 
pion multiplicities with respect to the data: The calculated average pion
multiplicity $< n_\pi> \simeq 4.5$ compared with the experimental
value of $\simeq 5.0$. We would like to remark, that non-vanishing 
contribution of the $n_\pi=2$ channel in calculations is completely due to the 
final states with other particles: $\pi \pi \eta$ (78\%), $\pi \pi K \bar K$ (14\%), 
$\pi \pi \eta \eta$ (6\%) and $\pi \pi$ + photons (2\% before $\eta$ decay). 
In calculations, we took into account $\eta$ decays into $2\gamma$ or into final states 
with pions and disregarded photons afterwards.
However, it is not clear to us how photons were counted in the data (see also
Ref. \cite{Vandermeulen88}).

\begin{figure}

\includegraphics[bb = 106 150 589 503]{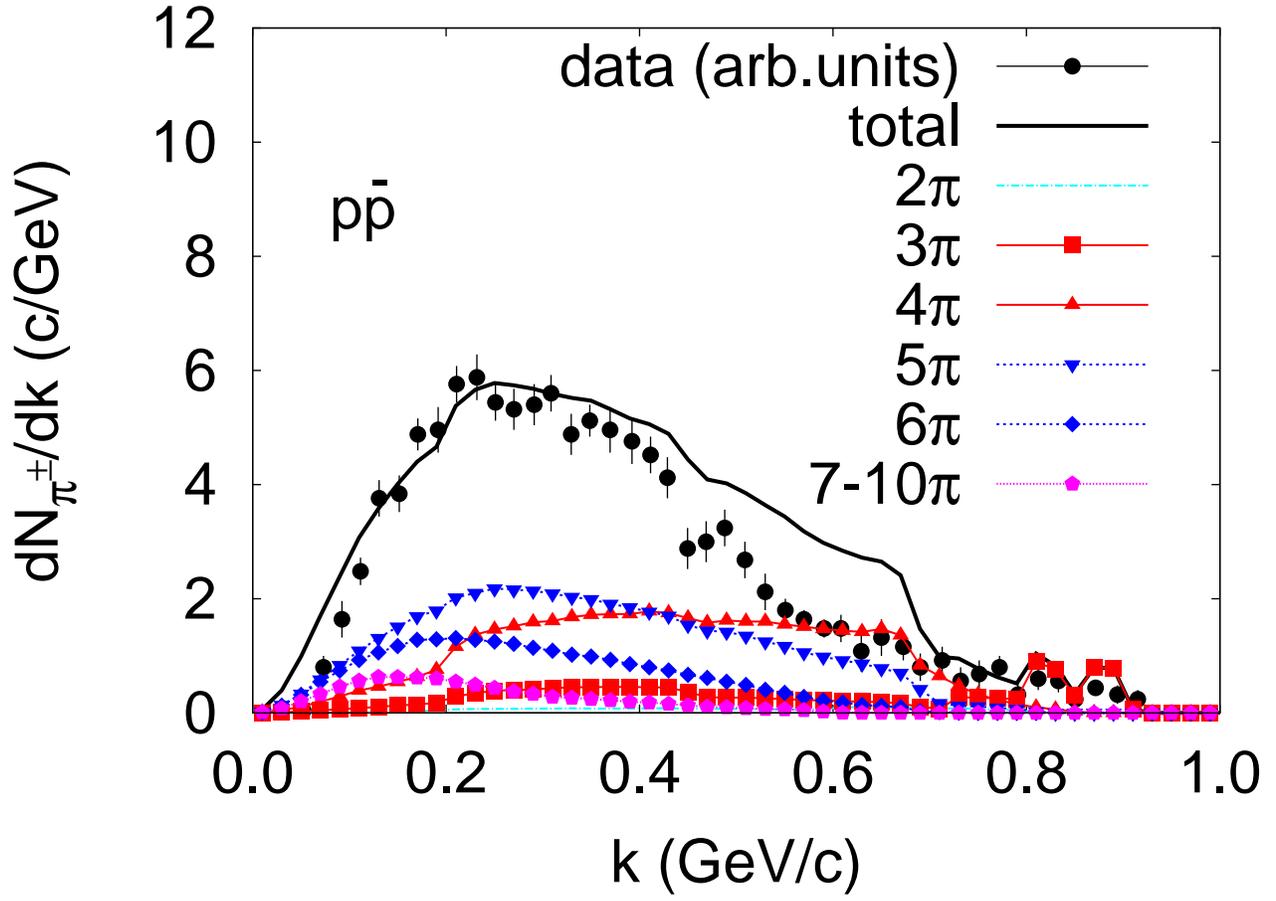}

\vspace*{1cm}

\caption{\label{fig:pimom_ppbar} (color online) Charged pion momentum distribution
for $p \bar p$ annihilation at rest in vacuum. The total calculated distribution 
is shown by the thick solid line. The calculated partial contributions from events with 
various pion numbers are also depicted (see key for notations). The calculations are
normalized to the number of charged pions per annihilation event. Data from Ref. \cite{Dover92} 
are in arbitrary units and are rescaled to agree with calculations at $k=0.3$ GeV/c.}

\end{figure}

Fig.~\ref{fig:pimom_ppbar} shows the calculated charged pion momentum distributions
in the c.m. frame of the annihilating $p \bar p$ pair at rest in vacuum. 
From the partial contributions of the channels with various pion multiplicities
we observe, as expected, that the hard(soft) part of the total momentum distribution 
is populated mainly by the low(high) pion multiplicity events. 
The experimental data are described reasonably well, except for the
momenta $0.5 \leq k \leq 0.7$ GeV/c, where the calculations significantly
overestimate the data. 

We believe that the accuracy of the model in describing the data in 
Figs.~\ref{fig:pimul_ppbar} and \ref{fig:pimom_ppbar} is sufficient for the 
exploratory studies of global observables in the present work. 
Certainly, the improvement of the annihilation model is needed 
to perform more detailed study of the mesonic final states in the 
annihilation. Below we concentrate more on the in-medium effects on
the annihilation. 

Due to the mean field, the invariant energy of the annihilating $\bar p N$ pair can 
be substantially below the vacuum threshold value of $2m_N$. This makes the direct application 
of the JETSET  model for the $\bar p$-annihilation in nuclei physically and numerically 
problematic. To overcome this difficulty, we introduced the corrected invariant energy
as follows \cite{LBGM07,WLM05}:
\begin{equation}
        \sqrt{s_{\rm corr}} = \sqrt{s^\star} - 2(m_N^\star-m_N)~,            
                                                                                 \label{srtsCorr}
\end{equation}
where $s^\star = (p_{\bar p}^\star + p_N^\star)^2$. The quantity $\sqrt{s_{\rm corr}}$ is
a vacuum analog of the total in-medium invariant c.m. energy $\sqrt{s}$ 
with $s=(p_{\bar p} + p_N)^2$. Provided that $\sqrt{s} > 4m_\pi$, we have used $\sqrt{s_{\rm corr}}$ 
in the JETSET simulation in order to produce the mesonic final states. This lower limit
of $\sqrt{s}$ is due to the fact that the JETSET model does not generate
enough direct $2\pi$ and $3\pi$ annihilation final states. 

In order to take into account the in-medium effects, in-particular, to ensure the correct
in-medium threshold condition $\sqrt{s} > m_1 + m_2 + ... +m_{n_{\rm mes}}$,
where $m_1,~m_2,~...,m_{n_{\rm mes}}$ are the vacuum masses of the produced mesons,
the annihilation event was accepted with the probability
\begin{equation}
   {\cal P} = \frac{ \Phi_{n_{\rm mes}}(\sqrt{s};m_1,m_2,...,m_{n_{\rm mes}}) }%
              { \Phi_{n_{\rm mes}}(\sqrt{s_{\rm corr}};m_1,m_2,...,m_{n_{\rm mes}}) }~,     \label{P}
\end{equation}
where
\begin{eqnarray}
   \Phi_{n_{\rm mes}}(\sqrt{s};m_1,m_2,...,m_{n_{\rm mes}})
    & = & \int\,\frac{ d^3 k_1 }{ (2\pi)^3 2\omega_1 } 
      \int\,\frac{ d^3 k_2 }{ (2\pi)^3 2\omega_2 } \cdots 
      \int\,\frac{ d^3 k_{n_{\rm mes}} }{ (2\pi)^3 2\omega_{n_{\rm mes}} }     \nonumber \\  
    & \times &  \delta^{(4)}(p_{\bar p} + p_N  - k_1 - k_2 - ... - k_{n_{\rm mes}} )        \label{Phi_nmes}
\end{eqnarray}
is the invariant phase space volume, $k_i=(\omega_i,{\bf k_i})$ are
the four-momenta of the produced mesons satisfying the vacuum mass shell conditions
$m_i^2=k_i^2$, $i=1,2,...,n_{\rm mes}$. Finally, the three-momenta of the produced
mesons in the c.m. frame of the annihilating $\bar p N$ pair were multiplied by
the common factor adjusted to get the correct in-medium total c.m. energy $\sqrt{s}$.

In-fact, the way we simulate the in-medium effects Eq.(\ref{P}) implies using 
the vacuum matrix elements of the annihilation channels, which are given by
the JETSET model, while taking into account the in-medium effects in the phase
space factors only. Similar procedures have been applied earlier in Refs. 
\cite{Mish05,LBGM07,WLM05}.

At $2m_\pi < \sqrt{s} \leq 4m_\pi$, the final $2\pi$ or $3\pi$ channel was chosen by Monte-Carlo 
according to the probability ratio
\begin{equation}
   \frac{ {\cal P}_{2\pi} }{ {\cal P}_{3\pi} } = R_0
   \frac{ \Phi_2(\sqrt{s};m_\pi,m_\pi) \Phi_3(2m_N;m_\pi,m_\pi,m_\pi) }%
        { \Phi_2(2m_N;m_\pi,m_\pi) \Phi_3(\sqrt{s};m_\pi,m_\pi,m_\pi) }~,          \label{Ratio}
\end{equation}
where $R_0=0.152$ is the ratio of the $2\pi$ and $3\pi$ final state probabilities for
the $p \bar p$ annihilation at rest (see Table VI in Ref. \cite{Mish05}). For the zero total charge
$Q$ of the annihilating $\bar p N$ pair, the charge states of the outgoing pions  were also
determined from the data compilation of Ref. \cite{Mish05}. Since for $Q=\pm 1$ the data are absent,
the charges of the $3\pi$ final states were determined by assuming that the $\pi^Q \pi^0 \pi^0$
and $\pi^Q \pi^+ \pi^-$ final channels have equal probabilities. The momenta of the outgoing
pions were distributed microcanonically according to the available two- or three-body
phase space. 

After the annihilation is simulated, the residual nucleons and produced mesons
were propagated in time according to the full kinetic equations (\ref{kinEq}),
including both the baryonic mean fields and collision integrals. This takes into account
the entropy production caused by the two-body collisions at the expansion stage. 
Moreover, important processes of the meson rescattering
and absorption, e.g. $\pi N \to \Delta \to \pi N$ or $\pi N \to \Delta$,
$\Delta N \to N N$ are included in the collision integral. These processes influence 
the observed particle spectra.

\section{ Time evolution of bound $\bar p$-nuclear systems }

\subsection{Initial compression stage}

As demonstrated in Refs. \cite{Buer02,Mish05} by static RMF calculations, 
a deeply-bound antiproton-nucleus system can be significantly compressed
as compared with a normal nucleus. Now we want to study the real
dynamics of such a system starting from the unperturbed nuclear ground 
state at $t=0$.  

\begin{figure}

\includegraphics[bb = 64 50 570 705, scale = 0.9]{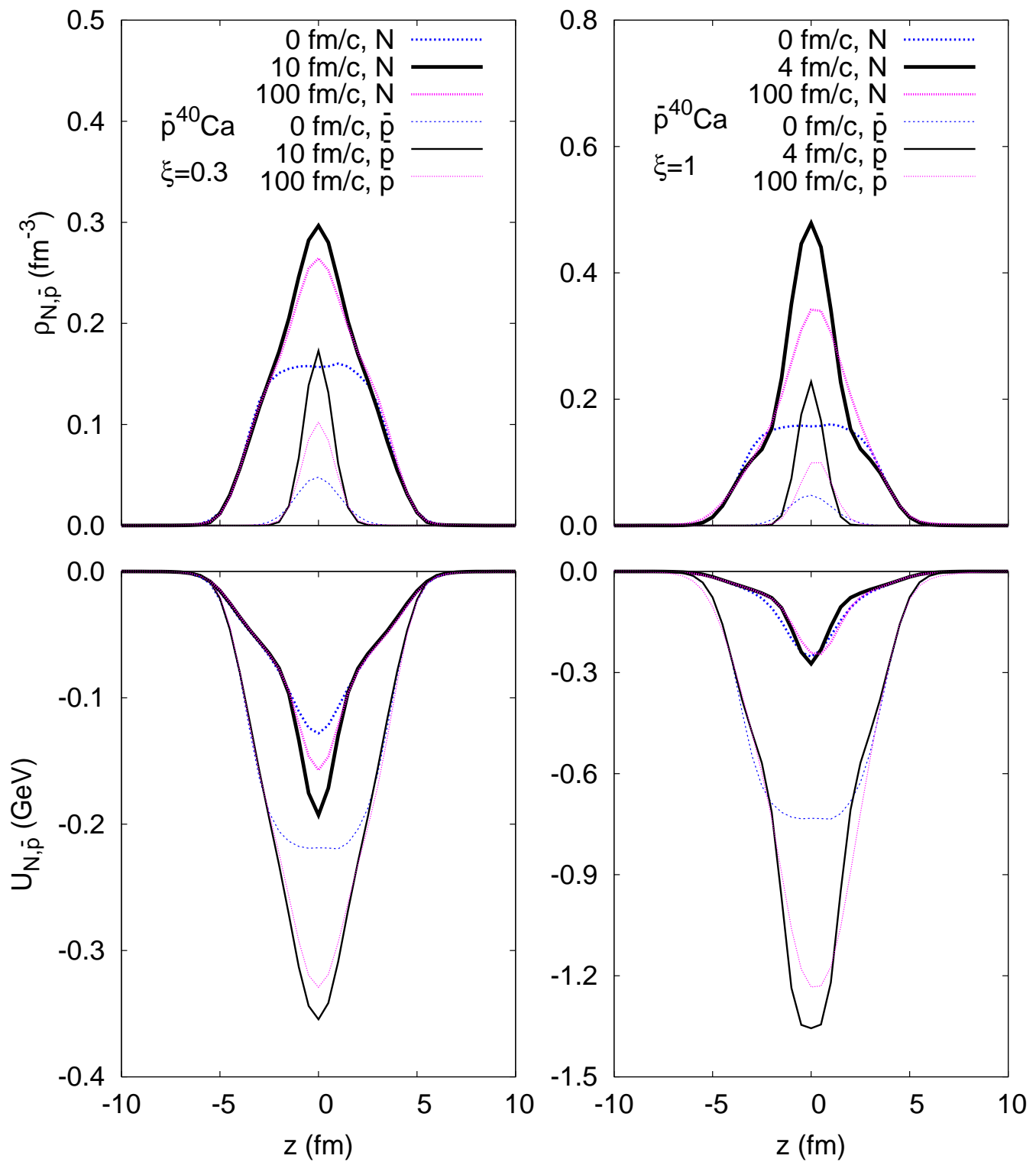}

\vspace*{1cm}

\caption{\label{fig:rho_U_vs_z} (color online) Nucleon and antiproton densities
(top panels) and potentials (bottom panels) vs coordinate $z$ on the axis passing
through the center of the $\bar p$$^{40}$Ca system at selected times indicated
in the figure.
The calculations with the scaling factor $\xi=0.3$ ($\xi=1$) are shown in the left (right)
panels. Please, notice different scales on vertical axis.}
   
\end{figure}

Fig.~\ref{fig:rho_U_vs_z} (top panels) shows the nucleon and antiproton
density profiles calculated at different times along the axis $z$ drawn through 
the center of the $\bar p$$^{40}$Ca system. Fig.~\ref{fig:rho_U_vs_z} (bottom
panels) also shows the nucleon and antiproton potentials 
$U_j \equiv g_{\omega j} \omega^0 + g_{\sigma j} \sigma$, $j=N,~\bar p$,
along the same axis. Left and right panels present results for $\xi=0.3$
and $\xi=1$, respectively. We see that the initial configuration is unstable 
and the system starts to shrink. Both nucleon and antiproton central 
densities grow quite fast, reaching their maxima within several fm/c. In the course 
of the compression process, the nucleon potential becomes deeper in the case 
of the reduced antiproton coupling constants ($\xi=0.3$) and does not, practically, 
change in the $G$-parity motivated case ($\xi=1$). The antiproton potential deepens 
quite strongly with time for the both sets of the antiproton coupling constants.

In the case of $\xi=0.3$, the first maximum of the central nucleon density 
($\rho_N = 0.30$ fm$^{-3}$) is reached at $t=10$ fm/c. At later time the
system rebounds and oscillates approaching gradually a static configuration with 
the nucleon density $\rho_N \simeq 0.26$ fm$^{-3}$ at the center. 
Since the annihilation is switched off in this calculation, the compressed configuration
may exist, in principle, infinitely long time. However, due to numerical
reasons, the stability is destroyed by a gradual test particle escape from a box in 
the coordinate space, where the mean field is computed. Nevertheless, the numerical 
accuracy is good enough to trace the stable system up to at least $t=100$ fm/c.

In the case of $\xi=1$, the compression process is much faster than in the case
of $\xi=0.3$. Already at $t=4$ fm/c we observe the first maximum of the central
nucleon density with $\rho_N=0.48$ fm$^{-3}$. A smaller value 
$\rho_N \simeq 0.34$ fm$^{-3}$ is reached asymptotically after some 
oscillations.

\begin{figure}

\includegraphics[bb = 64 50 570 705, scale = 0.9]{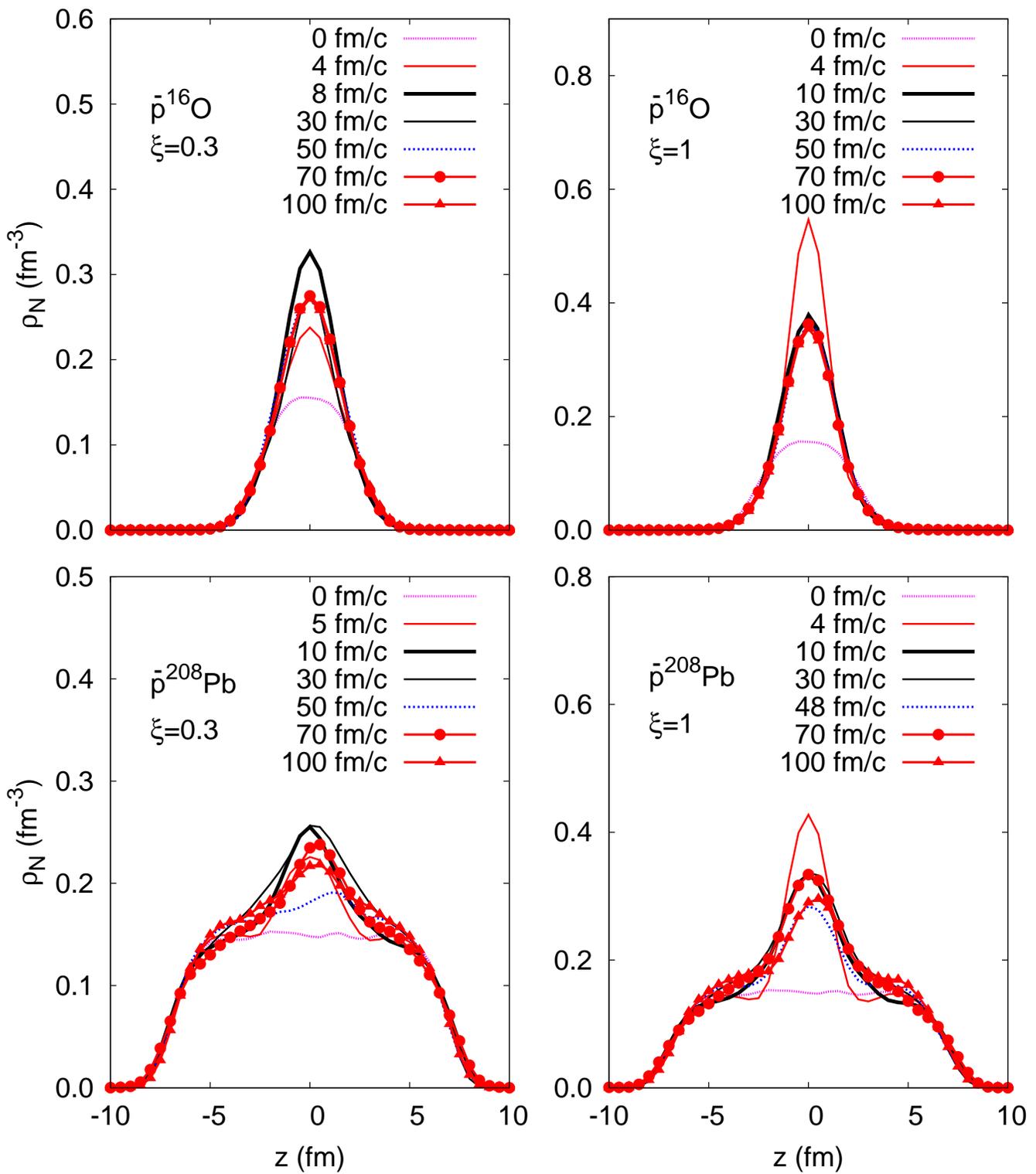}

\vspace*{1cm}

\caption{\label{fig:rho_vs_z} (color online) Nucleon densities along coordinate $z$
at various time moments for $\bar p$$^{16}$O and $\bar p$$^{208}$Pb at $\xi=0.3$ and $\xi=1$.}
\end{figure}

In Fig.~\ref{fig:rho_vs_z} we compare time evolution of the nucleon density distribution 
along the central axis $z$ for the light ($\bar p$$^{16}$O) and heavy ($\bar p$$^{208}$Pb)
systems. For $\bar p$$^{16}$O, the bell-like shape of the density distribution is reached
quite fast. However, in the case of $\bar p$$^{208}$Pb, we observe a quickly growing
peak in the center, while peripheral nucleons still do not react on the compression.
This leads to the delayed shape equilibration via a complicated compression-decompression
cycle.

\begin{figure}

\includegraphics[bb = 64 75 570 700, scale = 0.9]{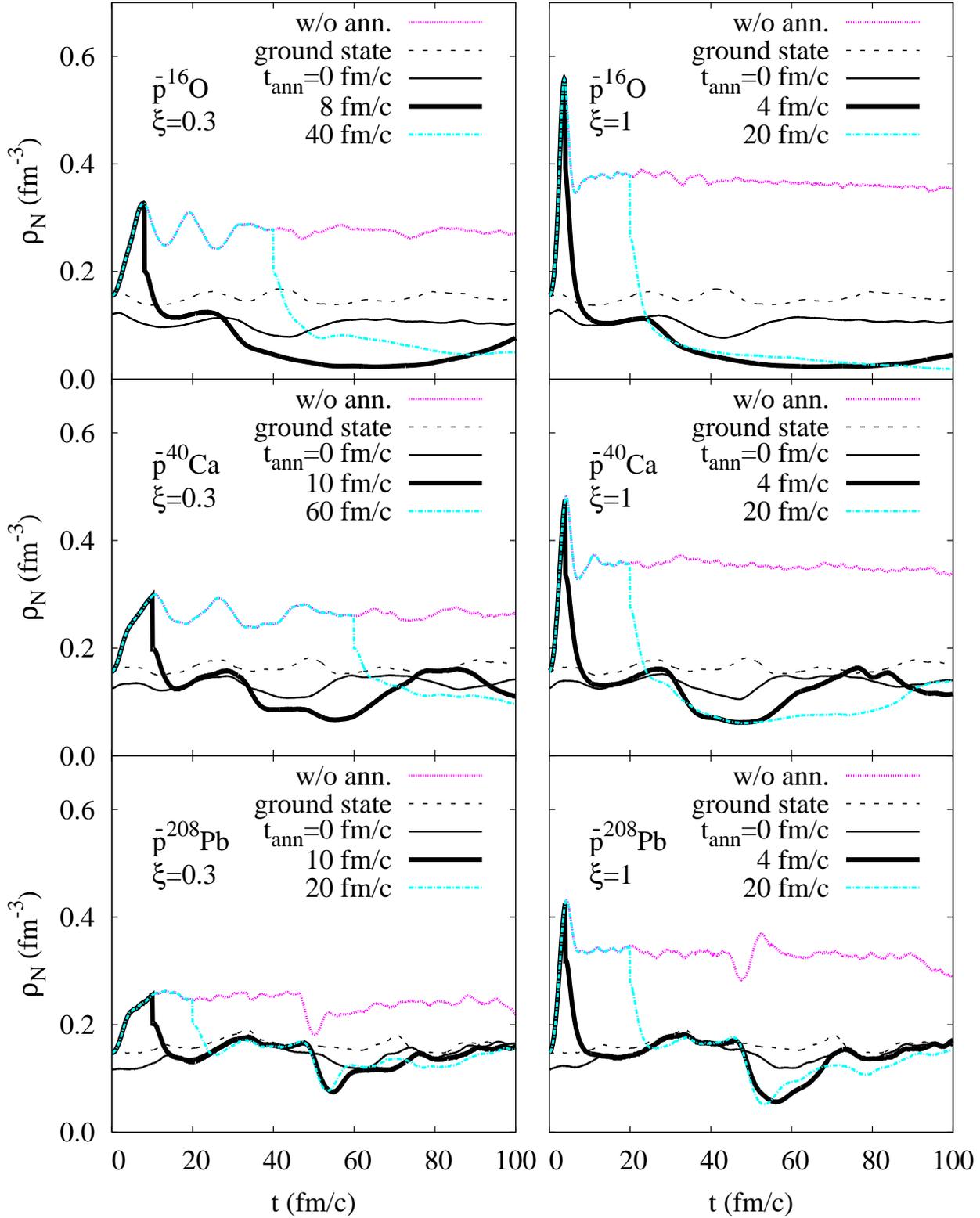}

\vspace*{0.5cm}

\caption{\label{fig:rho_vs_t} (color online) Time dependence of the nucleon density
at the center of the $\bar p$$^{16}$O, $\bar p$$^{40}$Ca and $\bar p$$^{208}$Pb 
systems for two values of the scaling factor $\xi=0.3$ (left panels) and $\xi=1$ 
(right panels) of the antiproton coupling constants. The dotted line shows 
the calculation without $\bar p$ annihilation. The thin solid, thick solid and dash-dotted 
lines show the results with annihilation simulated at various times $t_{\rm ann}$
indicated in the figure. The dashed lines show the central nucleon density evolution for 
the corresponding ground state nucleus without an antiproton.}
   
\end{figure}

Fig.~\ref{fig:rho_vs_t} shows the time evolution of the central
nucleon density for the three systems: $\bar p$$^{16}$O,
$\bar p$$^{40}$Ca and $\bar p$$^{208}$Pb. The case without annihilation
is shown by the dotted lines. For a comparison, we also present the central density
time evolution in the respective ground state nuclei without an 
antiproton inside (dashed lines). We see, that at long enough
times of the order of several tens fm/c, the static compressed configuration 
is indeed reached. The small oscillations of the central density in
the compressed system visible at $t > 50$ fm/c are approximately of 
the same amplitude as the oscillations of the respective ground state.
Thus, the reason for these small amplitude oscillations is the
fermionic ground state instability due to the classical treatment of particles 
\cite{Bondorf97,Wilets77,Dorso87,Peilert91,Bonasera01}. In principle, this  
instability can be removed by either employing the Pauli potential as in   
Refs. \cite{Wilets77,Dorso87,Peilert91} or by adding a friction
force to the Hamiltonian equations of motions for the test particles
\cite{Bonasera01}. Such modifications are beyond the scope of the present 
work. We expect that they would not essentially 
modify the density profiles of the compressed configurations, which are 
in an overall agreement with the previous static Hartree calculations of 
Refs. \cite{Buer02,Mish05}.

One can also notice (see lower panels of Fig.~\ref{fig:rho_vs_t}) a peculiar 
feature of the $\bar p$$^{208}$Pb system: the dip in the central nucleon density 
at $t\simeq50$ fm/c. This is mostly a consequence of the delayed shape equilibration 
mentioned above in discussing Fig.~\ref{fig:rho_vs_z}. The especially strong density 
drop for $\xi=0.3$ is partly caused by the symmetry loss due to the finite number of 
test particles. The dip is absent in lighter systems, since the shape 
equilibration for them is much faster ($10\div20$ fm/c, see upper panels in 
Figs.~\ref{fig:rho_U_vs_z} and \ref{fig:rho_vs_z}).

Since the compression time is of primary importance, we have also studied
the sensitivity of our results to the width $\sigma_r$ of
the initial antiproton DF (\ref{cohState}). We have found,
that for a larger (smaller) width the compression time becomes
somewhat longer (shorter). In-particular, for the $\bar p$$^{16}$O system at
$\xi=1$ the time needed to reach the first density maximum is 5 fm/c (2.5 fm/c)
for $\sigma_r=2$ fm ($\sigma_r=0.5$ fm). For the same system at $\xi=0.3$ the
first density maximum is reached at 10 fm/c (3 fm/c) for $\sigma_r=2$ fm 
($\sigma_r=0.5$ fm).
The value of the maximum density reached in the compression process is practically
insensitive to $\sigma_r$.

We have to admit also, that there is some numerical uncertainty in our calculations 
due to the choice of the width $L$ of the test particle Gaussian and the grid step size.
E.g. in the calculation with $\Delta_x=\Delta_y=\Delta_z=L=0.33$ fm for the lightest
system $\bar p$$^{16}$O the maximum and saturation densities are $20$ \% 
higher than in calculation with $\Delta_x=\Delta_y=\Delta_z=L=0.5$ fm.
Setting the smaller grid step is not feasible for technical reasons.
Overall, this numerical uncertainty is comparable to the one due to different
choices of the scaling factor $\xi$ in our calculations.

\subsection{ Post-annihilation dynamics of residual nuclei }

Next, we study the dynamics of a residual nucleus after sudden
annihilation of an antiproton. The annihilation was simulated  
as described in Sect. IIE. For each considered $\bar p$A 
system and the scaling factor $\xi$, three different 
annihilation times $t_{\rm ann}$ have been chosen: They correspond to 
(i) the early ($t_{\rm ann}=0$) annihilation from a 
non-compressed ground-state nucleus, (ii) the annihilation at the time 
moment when the first maximum of the central density is reached, 
and (iii) the late annihilation from an asymptotic compressed 
configuration.

One can see from Fig.~\ref{fig:rho_vs_t}, that the annihilation from a 
non-compressed ground state nucleus (thin solid lines) does not lead to 
significant expansion of the residual nuclear system. The central nucleon 
density stays always below but close to $\rho_0$ in this case. In the case of
annihilation from the compressed configurations (thick solid and 
dash-dotted lines), we observe that, for the light systems $\bar p$$^{16}$O 
and $\bar p$$^{40}$Ca, the central nucleon density decreases sharply after
annihilation and reaches values well below $\rho_0$. This is 
a clear indication of the collective expansion of a system from the initially 
compressed state. On the other hand, for the heavy 
$\bar p$$^{208}$Pb system, the expansion is not very pronounced at any choice 
of the annihilation time.

\begin{figure}

\includegraphics[scale = 0.9]{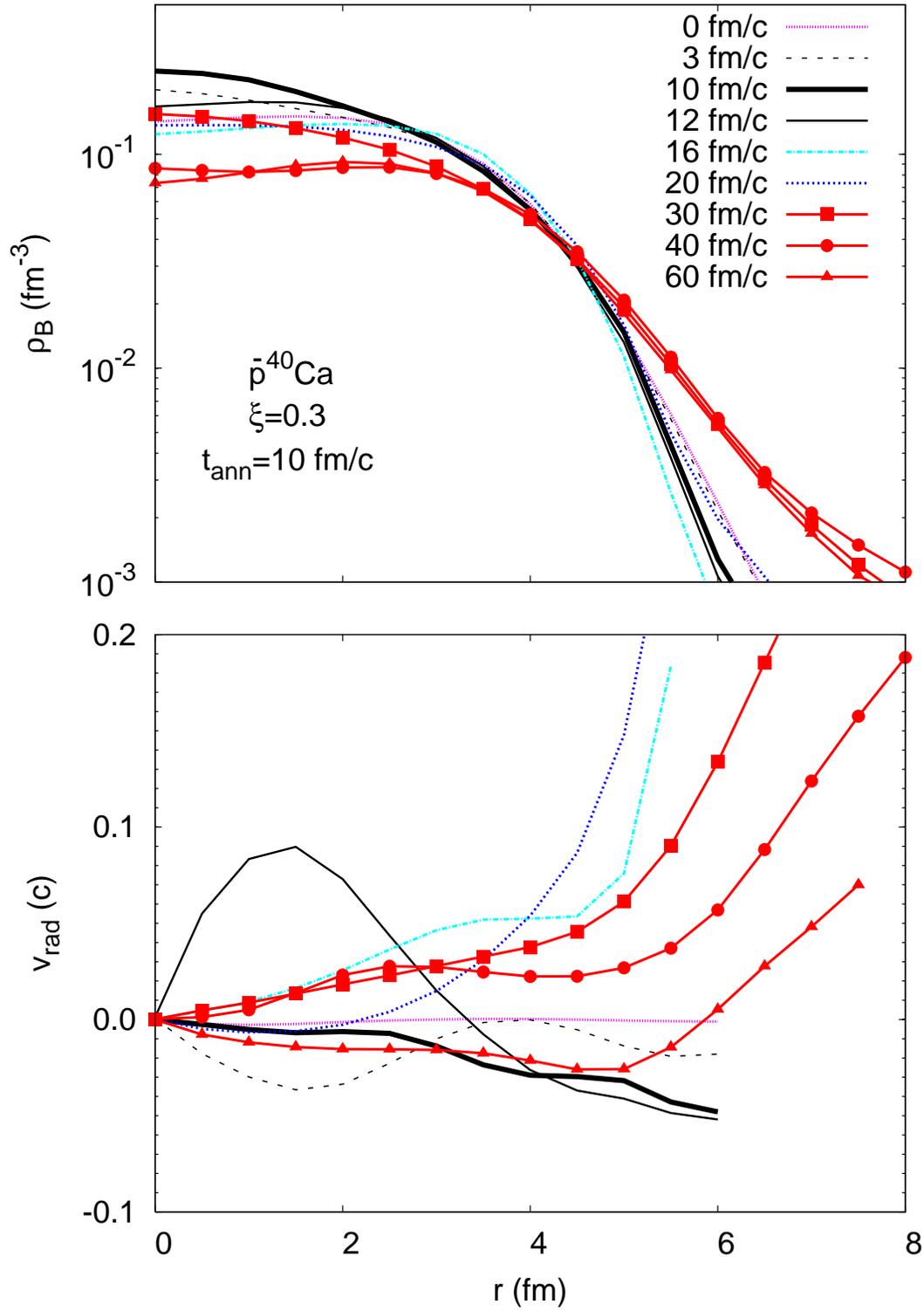}

\vspace*{0.5cm}

\caption{\label{fig:vrad_Ca40_xip03} (color online) The baryon density (top panel)
and the radial collective velocity (bottom panel) as functions
of the radial distance for the $\bar p$$^{40}$Ca system computed with $\xi=0.3$.
The annihilation time $t_{\rm ann}$ was set to 10 fm/c.}

\end{figure}

\begin{figure}

\includegraphics[scale = 0.9]{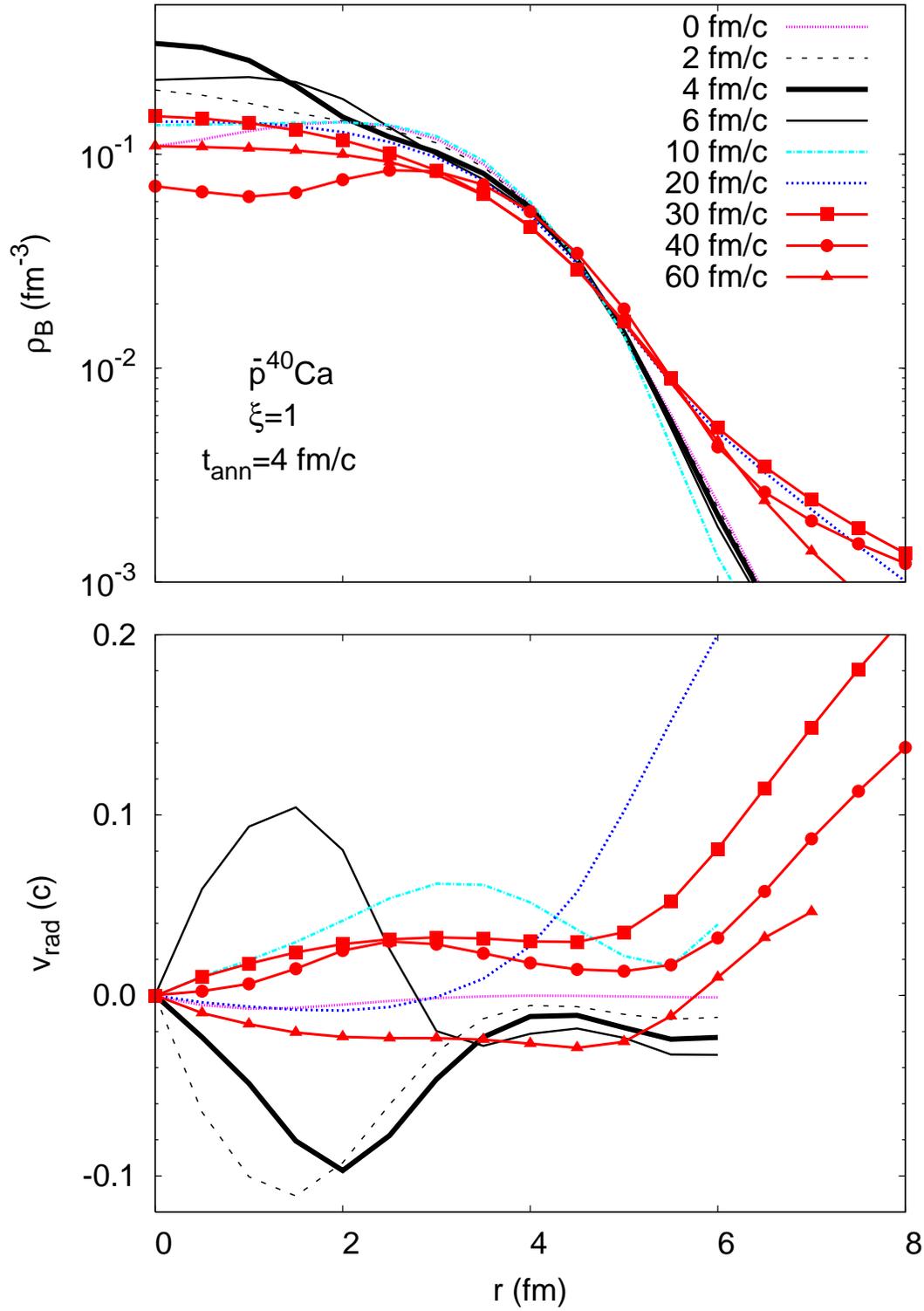}

\vspace*{0.5cm}

\caption{\label{fig:vrad_Ca40_xip1} (color online) Same as in 
Fig.~\ref{fig:vrad_Ca40_xip03}, but for $\xi=1$ and $t_{\rm ann}=4$ fm/c.}

\end{figure}

It is interesting, that if the annihilation is switched on at the first
density maximum (thick solid lines), then after an abrupt falling down 
the central density stays for some time $\sim 10-20$ fm/c close to $\rho_0$ 
before decreasing further.
This is explained by an inertial compression:  
After annihilation, the periphery of a residual nucleus still continues to move 
to the center during some time until rebound. In the case of the $\bar p$$^{40}$Ca 
system, this is demonstrated in Figs. \ref{fig:vrad_Ca40_xip03}
and \ref{fig:vrad_Ca40_xip1}, where we show the baryon density and the 
radial collective velocity, 
\begin{equation}
   v_{\rm rad}= {\bf r}\cdot {\bf v}_{\rm coll}/r,       \label{vrad}
\end{equation}
at several times as a function of the radius $r$.
The collective velocity has been determined as
\begin{equation}
   v_{\rm coll}^\alpha = T^{\alpha0}/T^{00}~,~~~\alpha=1,2,3.  \label{vcoll}
\end{equation}
At $r >4$ fm, the evolution of the radial collective velocity field is noticeably
influenced by the fermionic ground state instability discussed above. Nevertheless,
one can still observe the inertial compression. Indeed, the radial collective 
velocity at $t=12$ fm/c for $\xi=0.3$ and at $t=6$ fm/c for $\xi=1$ reveals
the fast outward motion at the center, while the peripheral nucleons still
continue to move to the center. This explains plateaus in the central density 
evolution near $t\simeq20$ fm/c in Fig. \ref{fig:rho_vs_t}.  
At $t \simeq 30$ fm/c the whole system starts to expand. This 
is reflected in the monotonically increasing radial collective velocity with radius. 
The especially strong rise of $v_{\rm rad}$ 
at large $r$ is due to emission of fast particles. At later times $t \simeq 50\div60$ 
fm/c the expansion is replaced by the inward motion of the matter 
in the central zone. However, the fast particles are still continuing to escape 
from the dense region. We expect that in reality the system will break-up into 
fragments before the inward motion will start (see discussion Sect. IV A).

\section{ Observable signals }

\subsection{ Multifragmentation of residual nuclei }

It is presently well established (see \cite{CCR04} and refs. therein)
that nuclear matter at low densities ($\rho < 0.6\,\rho_0$) becomes unstable
with respect to small density perturbations, so called spinodal instability. 
However, in order these density perturbations
to develop into nuclear fragments, the system must stay long enough time 
$\sim 30$ fm/c in the spinodal region. One can see from Fig.~\ref{fig:rho_vs_t}, 
that the light systems $\bar p$$^{16}$O and $\bar p$$^{40}$Ca
spend a long time in this region. 
Therefore, the residual nuclear systems can undergo a multifragment break-up, 
if the annihilation happens in the compressed configurations. In other words,
the multifragment break-up of nuclei after the $\bar p$-annihilation may serve
as a signal of the compression prior the annihilation.

In Table \ref{tab:fragmenting_source} we collect the estimated parameters 
of fragmenting sources for the $\bar p$$^{16}$O and $\bar p$$^{40}$Ca
systems. The case of $\bar p$-annihilation from the state of maximum
central density is considered here (see thick solid lines in 
Fig.~\ref{fig:rho_vs_t}).
The sources have been determined by selecting nucleons in 
the space region where the baryon density is larger than $\rho_{\rm min}=0.01\rho_0$.
They are characterized by the neutron ($N$) and proton ($Z$) numbers, the collective 
kinetic energy per nucleon ($E_{\rm kin}^{\rm coll}$) and the residual excitation energy 
per nucleon ($E_{\rm res}^\star$). These parameters are defined as
\begin{eqnarray}
   & & N = \int\limits_{\rho > \rho_{\rm min}}d^3r \rho_n({\bf r})~,~~~
       Z = \int\limits_{\rho > \rho_{\rm min}}d^3r \rho_p({\bf r})~,   \label{NZ} \\
   & & E_{\rm kin}^{\rm coll} = \frac{1}{A} \int\limits_{\rho > \rho_{\rm min}}d^3r
   \left( T^{00} - \sqrt{T^{\mu0}T_\mu^0}\,\right)~,                           \label{E_kin^coll} \\
   & & E_{\rm res}^\star = \frac{1}{A} \int\limits_{\rho > \rho_{\rm min}}d^3r\, T^{00}
       - E_{\rm g.s.}(N,Z) - E_{\rm kin}^{\rm coll}~,     \label{E_res^coll}
\end{eqnarray}
where $\rho_n$ and $\rho_p$ are the neutron and proton densities, respectively;
$A=N+Z$; and $E_{\rm g.s.}(N,Z)$ is the ground state energy per nucleon of a nucleus with 
neutron number $N$ and proton number $Z$ computed within our model.
The collective kinetic energy (\ref{E_kin^coll}) is calculated neglecting pressure effects.
Due to the initially isospin-symmetric nuclei and the neglect of the isovector mesons in the 
Lagrangian density (\ref{Lagr}), we obtained in all cases $N \simeq Z$ in the source. It turned 
out also, that the Coulomb interaction optionally included in some of the calculations does not 
change this result. Thus, only the charge numbers $Z$ are given in Table \ref{tab:fragmenting_source}. 
The time moment for determination of the source parameters has been chosen at $35\div40$ fm/c, 
when the central nucleon density is about $1/3\div1/2\rho_0$, i.e. inside the spinodal region. 
The calculated residual excitation energy is typically $6\div10$ MeV/nucleon that corresponds
to the temperatures $4\div7$ MeV. According to the statistical multifragmentation model \cite{SMM},
in this energy domain the multifragment break-up is the dominating decay channel
of residual nuclei.
The collective kinetic energy is $1.4\div2.1$ MeV/nucleon in the case of $\bar p$$^{16}$O.
This is well above the Coulomb energy of the source, which is only about 
0.4 MeV/nucleon for $N=Z=5$. Unfortunately, such a source is too small to experience 
the real multifragment break-up, rather a Fermi break-up into small clusters \cite{SMM}.
In the case of larger sources, produced in $\bar p$$^{40}$Ca 
annihilation the collective kinetic energy is considerably smaller, $0.3\div0.6$ MeV/nucleon, 
but still significant with respect to the total Coulomb energy $\simeq 1.0$ 
MeV/nucleon for $N=Z=16$. Thus, we expect some signs of collective expansion to be visible 
in kinetic energy spectra of produced fragments.  

\subsection{ Knock-out nucleon spectra }

Let us now consider other observable effects. Fig.~\ref{fig:Ekin_spectra} shows 
the c.m. kinetic energy spectra of the nucleons emitted from the $\bar p$$^{16}$O, 
$\bar p$$^{40}$Ca and $\bar p$$^{208}$Pb systems after the $\bar p$-annihilation.
In order to separate emitted nucleons from the bound nucleons of a residual
nucleus, we used a simple criterion: only those nucleons, both protons and neutrons, were
included in the spectra which are separated by at least 3 fm from the other 
test particles of a given parallel ensemble at $t=100$ fm/c.
One can see, that nucleons with the kinetic energy $E_{\rm kin} \gg E_{\rm F}$,
where $E_{\rm F}\simeq35$ MeV is the Fermi energy of the 
nuclear matter at $\rho_0$, are abundantly emitted.
Such nucleons are knocked-out from the nucleus by the mesons produced
after the annihilation \cite{Cahay82}. 

In Table \ref{tab:slopes} we list the slope parameters $T_N$ of the nucleon
kinetic energy spectra obtained by the Maxwell-Boltzmann fit
\begin{equation}
   \frac{d N_{\rm nuc}}{d E_{\rm kin}} 
          \propto \sqrt{E_{\rm kin}} \exp(-E_{\rm kin}/T_N)    \label{fit_nuc}
\end{equation}
in the region of $E_{\rm kin}=200\div500$ MeV. We would like to mention, that
the authors of Ref. \cite{Cahay82} report the slope temperature of about 60 MeV
for the kinetic energy spectrum of the emitted  protons in the case of
$\bar p$$^{40}$Ca system, which is not so far from our results for the annihilation 
in the ground-state nucleus at $t_{\rm ann}=0$. 

We want to emphasize that the kinetic energy spectra of nucleons emitted after 
the $\bar p$-annihilation from the compressed $\bar p$$^{16}$O,
and $\bar p$$^{40}$Ca systems are significantly harder than the spectra of nucleons 
from the annihilation at $t_{\rm ann}=0$. 
This can be explained by two effects. First, the collective expansion 
of the outer shell will increase the slope temperature, typically,
by several MeV (see Table \ref{tab:fragmenting_source} and Figs.~\ref{fig:vrad_Ca40_xip03},
\ref{fig:vrad_Ca40_xip1}). Second, just after the annihilation the nucleon potential
at the center of a nucleus grows suddenly by $\sim 80\div300$ MeV. This creates an additional
push for the fast nucleons emitted from the nucleus.   
Although the hardening effect is most pronounced for the lightest system $\bar p$$^{16}$O,
it is also quite visible for $\bar p$$^{40}$Ca.
For the heaviest system $\bar p$$^{208}$Pb, we observe almost
identical high energy tails of the nucleon spectra
for the different annihilation times. The reason  is that the collective 
expansion is practically absent in this system (see Fig.~\ref{fig:rho_vs_t}).
Also, the yield of fast nucleons is reduced by their subsequent rescatterings
in the residual nucleus.

\subsection{ Mesonic observables }

\begin{figure}

\includegraphics[bb = 40 0 554 730, scale = 0.8]{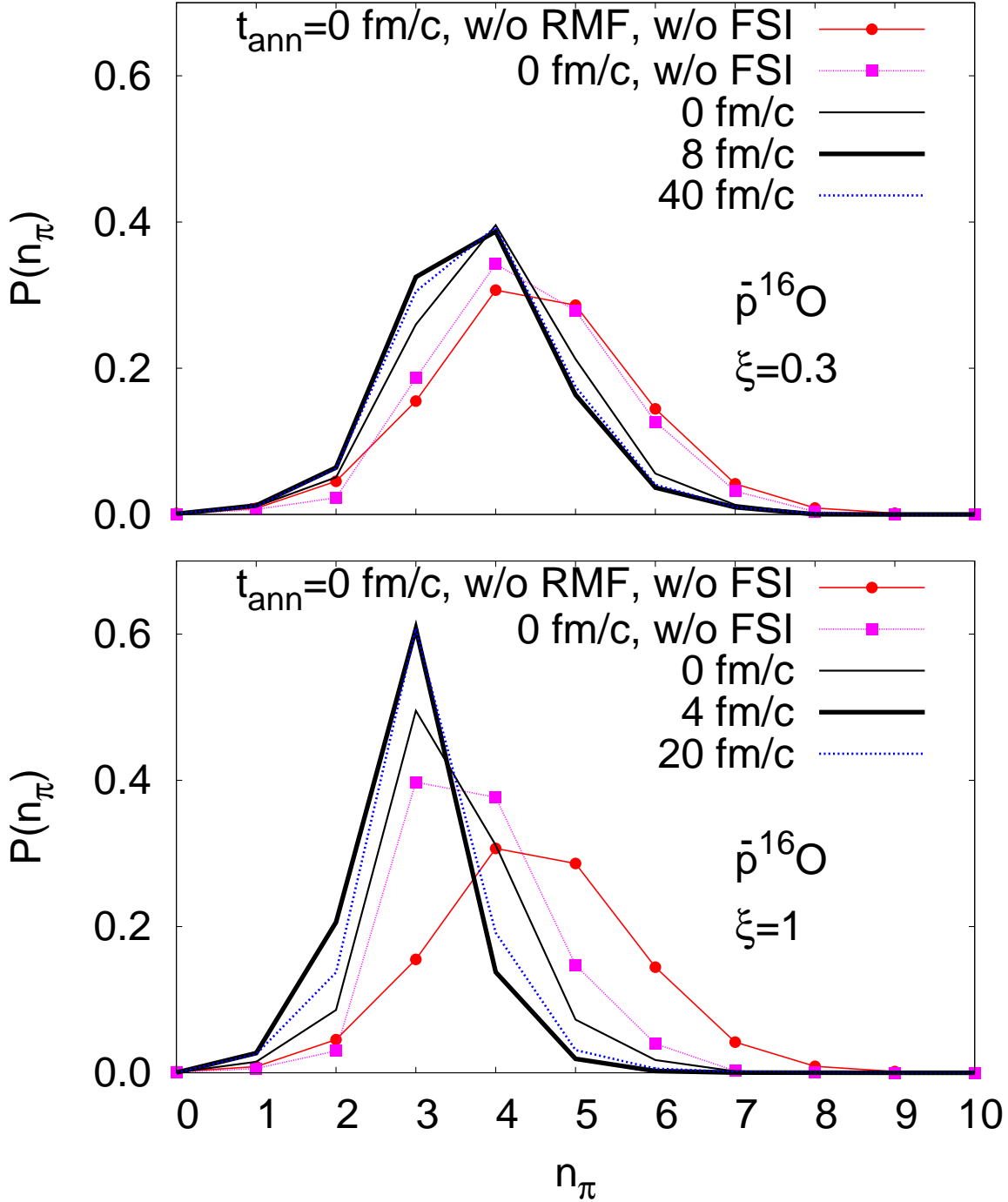}

\vspace*{0.5cm}

\caption{\label{fig:pimul_pbarO16} (color online) Pion multiplicity distributions
for the $\bar p$$^{16}$O system. The line with full circles shows the calculation
without mean field and without FSI after the annihilation.
The line with full boxes shows the result with mean field, but without FSI.
Other lines show the full calculation at various choices of the annihilation
time (shown in the key). The top (bottom) panel presents results for $\xi=0.3$
($\xi=1$).}

\end{figure}

\begin{figure}

\includegraphics[bb = 40 0 554 730, scale = 0.8]{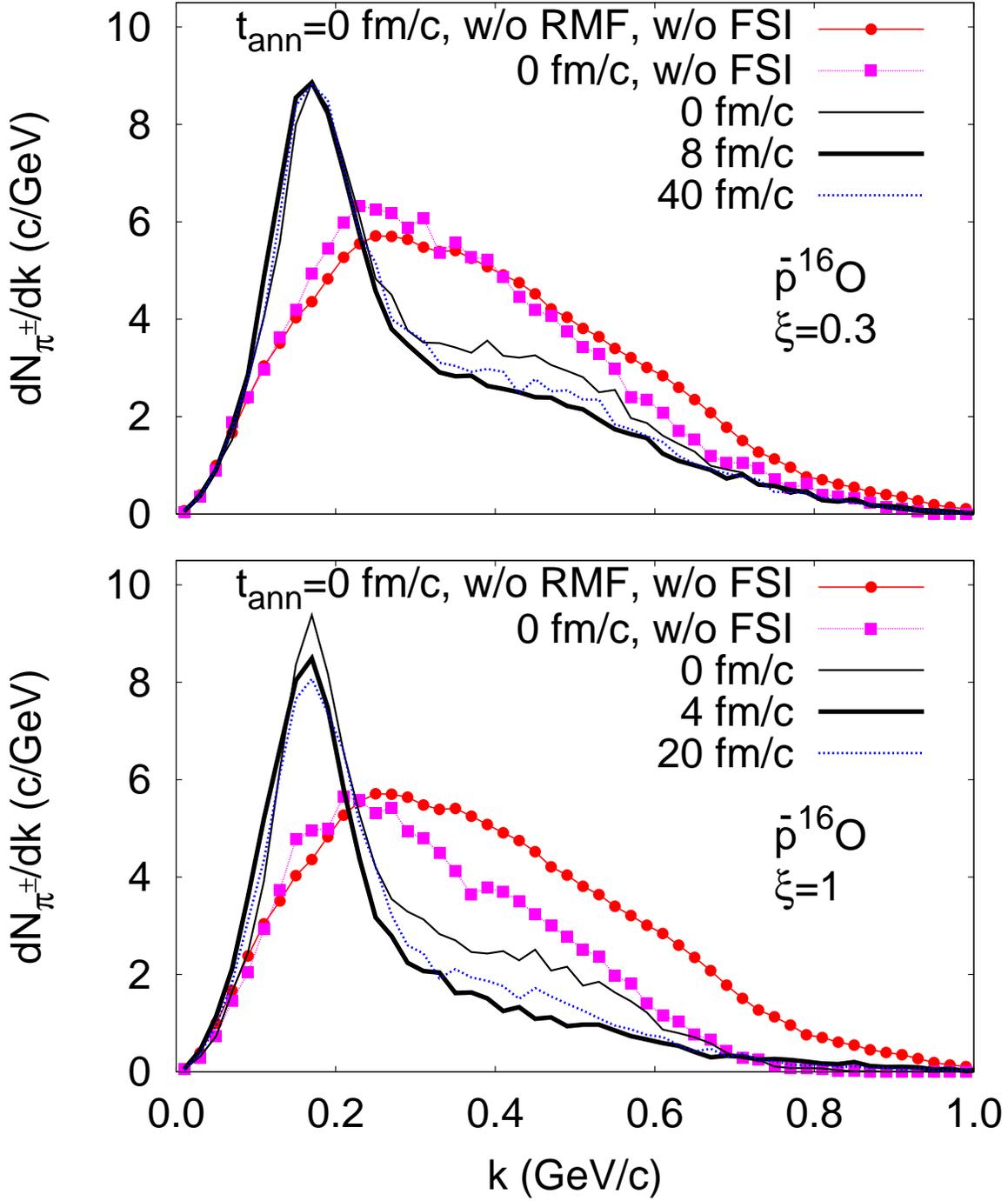}

\vspace*{1cm}

\caption{\label{fig:pimom_pbarO16} (color online) Same as in Fig.~\ref{fig:pimul_pbarO16},
but for the charged pion momentum distributions in the c.m. frame of the
$\bar p$$^{16}$O system.}

\end{figure}

\begin{figure}

\includegraphics[bb = 40 0 554 730, scale = 0.75]{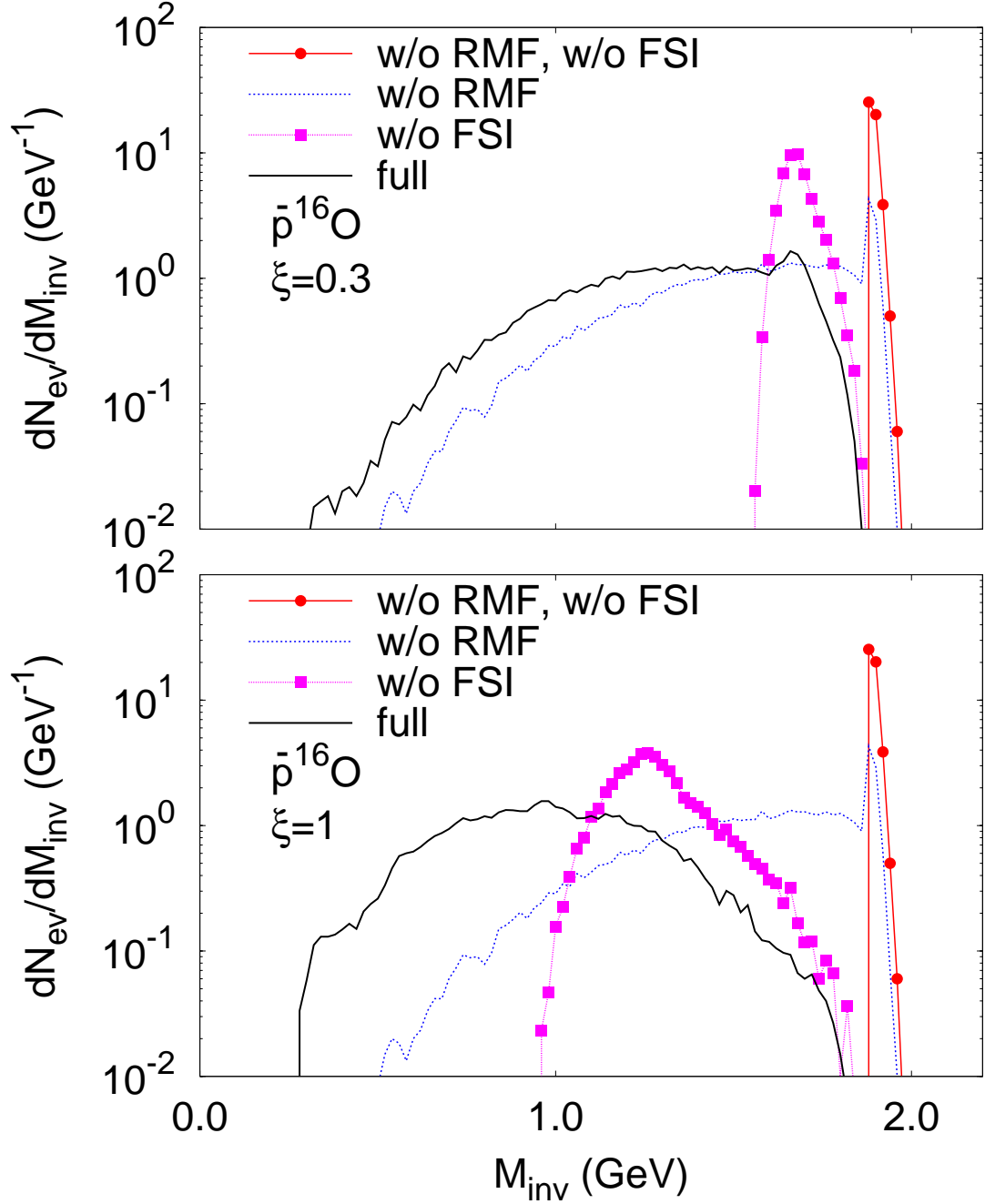}

\vspace*{0.5cm}

\caption{\label{fig:Minv_spectra_pbarO16} (color online) Distribution of the annihilation 
events on the total invariant mass of emitted mesons for the $\bar p$$^{16}$O system. 
The calculation without mean field and without FSI is shown by the line with full circles.
The results without mean field but with FSI are shown by the dotted line.
The line with full boxes shows the result with mean field, but without FSI.
The full calculation is presented by the solid line. Upper (lower) panel corresponds
to $\xi=0.3$ ($\xi=1$). Only calculations at $t_{\rm ann}=0$ are shown.}  

\end{figure}

The meson production from the $\bar p$-annihilation inside nucleus 
is influenced both by the mean field via the potentials of the annihilating
pair and by the final state interactions (FSI), i.e.
the two-body collisions and resonance decays. It is instructive to disentangle 
the contributions of the mean field effects from the rescattering 
and absorption effects. To this aim, we have performed additional calculations
by subsequently switching off the FSI and the mean field.
Corresponding results are shown in Figs.~\ref{fig:pimul_pbarO16},\ref{fig:pimom_pbarO16} 
and \ref{fig:Minv_spectra_pbarO16}.

Figure~\ref{fig:pimul_pbarO16} shows pion multiplicity distributions
for the $\bar p$$^{16}$O system. In the case of reduced antiproton coupling
constants ($\xi=0.3$), the mean field alone does not produce any noticeable
modification of the pion multiplicity distribution. On the other hand, FSI leads 
to a rather substantial shift toward smaller $n_\pi$. For the case of $\xi=1$, 
we observe a strong pion multiplicity reduction due to smaller $\sqrt{s}$ for
the $\bar p N$ annihilation, while the FSI effects are much weaker. 
One can also see a quite significant compressional effect for $\xi=1$, which is only 
very weak for $\xi=0.3$.

\begin{figure}

\includegraphics[bb = 120 75 626 730, scale = 0.9]{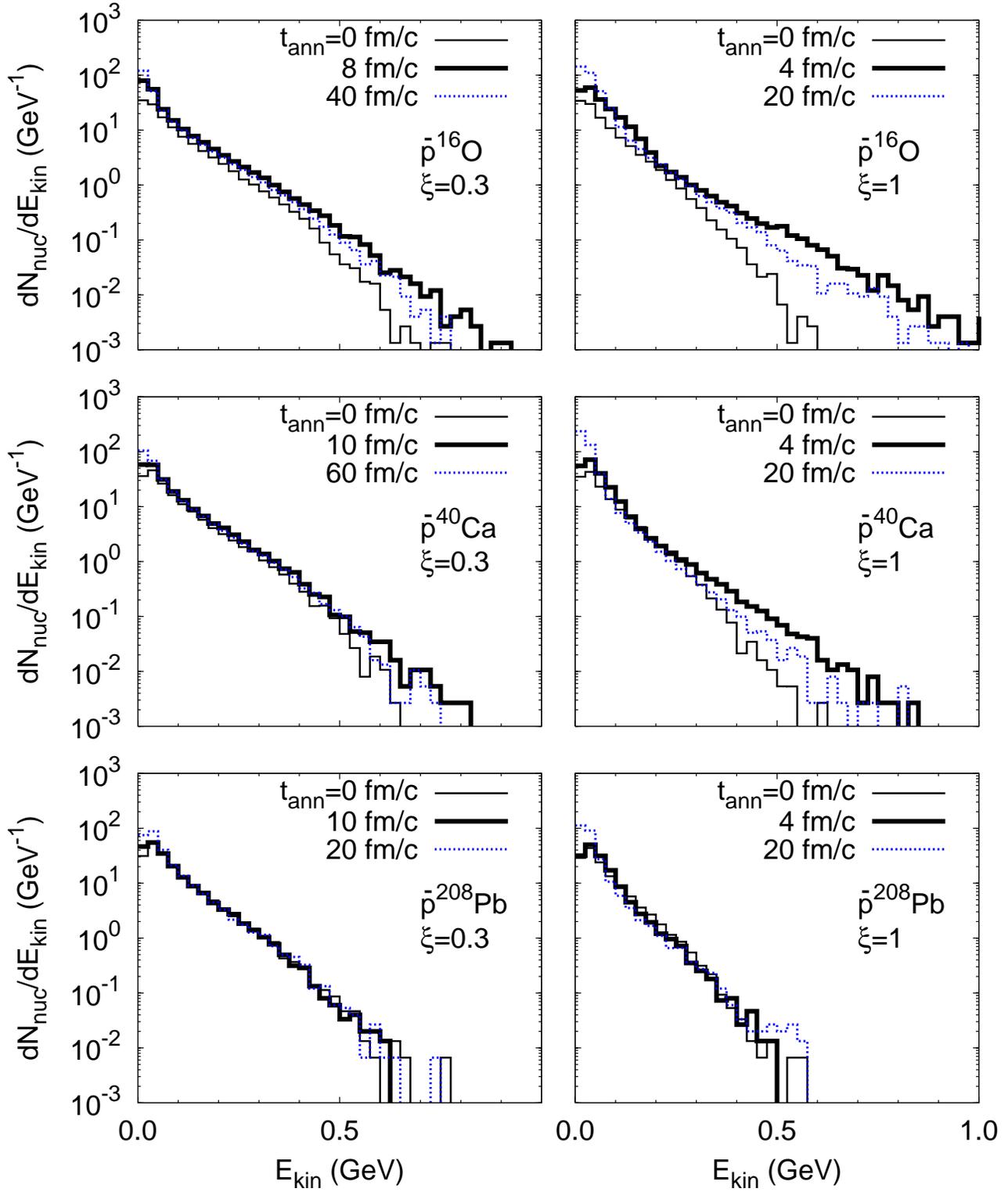}

\vspace*{0.5cm}

\caption{\label{fig:Ekin_spectra} (color online) Kinetic energy spectra of 
emitted nucleons in the c.m. frame for various $\bar p$A systems and values of the parameter
$\xi$. Different histograms correspond to different values of the annihilation time
$t_{\rm ann}$ indicated in the key.}

\end{figure}

In Fig.~\ref{fig:pimom_pbarO16}, we present charged pion momentum spectra 
in the c.m. frame of the $\bar p$$^{16}$O system. FSI strongly modifies these 
spectra, mostly due to the $\pi N \to \Delta \to \pi N$ 
processes, which effectively decelerate pions. As we have already observed earlier
in Fig.~\ref{fig:Ekin_spectra}, emitted nucleons gain energy, correspondingly.

The effect of the baryonic mean field on the pion momentum spectrum is relatively
moderate: we observe some depletion of the high-momentum tail, which is more
pronounced in the case of $\xi=1$. The compressional effect is visible in 
the reduction of the pion yield in the momentum range $0.3 \leq k \leq 0.6$ GeV/c.
This can be understood from Figs.~\ref{fig:pimom_ppbar} and 
\ref{fig:pimul_pbarO16}: in vacuum, the events with $n_\pi=5$ and $n_\pi=6$ contribute 
substantially to this momentum range, while the probability of such events is substantially
suppressed in a compressed nucleus due to the reduced annihilation phase space \cite{Mish05}.   

The pion momentum spectrum has clearly the two components: the slow pions, which
have undergone rescatterings via the $\Delta$-resonance excitation 
($k$ less than about 300 MeV/c), and the high energy pions, which were emitted 
from the system almost without secondary interactions. The similar result has been
obtained in earlier intranuclear cascade calculations \cite{Cahay82}.
Following Ref. \cite{Cahay82}, we have also fitted the low energy  part of
the pion spectrum ($E-m_\pi=100\div150$ MeV, $E=\sqrt{k^2+m_\pi^2}$) by the Maxwell-Boltzmann 
distribution   
\begin{equation}
   \frac{d N_{\pi^\pm}}{d k} = k^2 \exp(-E/T_\pi)~.    \label{fit_pion}
\end{equation}
This fit has produced the following slope temperatures of the charged pion momentum
spectrum for the $\bar p$$^{16}$O system: $T_\pi\simeq45$ MeV and 44 MeV for $\xi=0.3$
($T_\pi\simeq43$ MeV and 36 MeV for  $\xi=1$) in the case of early annihilation 
($t_{\rm ann}=0$) and annihilation at the time of maximum compression, respectively.
The extracted $T_\pi$ is smaller than
the slope temperature of $53$ MeV of the low energy pions for the $\bar p$$^{40}$Ca 
system reported in Ref. \cite{Cahay82}. However, in the calculation without mean field,
we obtain $T_\pi\simeq51$ MeV, which is in a good agreement with the result of Ref. 
\cite{Cahay82}.
Therefore, the difference between our results and those of Ref. \cite{Cahay82}
is caused by the mean field acting on the annihilating pair in medium. 
Moreover, we see the softening of the pion spectrum in the case of larger antinucleon couplings 
due to smaller $\sqrt{s}$ of the annihilating pair. 

Figure~\ref{fig:Minv_spectra_pbarO16} presents the distributions of the
annihilation events in the total invariant mass of produced mesons
from the $\bar p$$^{16}$O system. The invariant mass is defined as
\begin{equation}
   M_{\rm inv} = \left( ({\rm P}_{\rm mes}^0)^2 
                      - {\bf P}_{\rm mes}^2 \right)^{1/2}~,  \label{Minv}
\end{equation}
where ${\rm P}_{\rm mes}^\mu = \sum\limits_i p_i^\mu$ is
the sum of four-momenta of the mesons  produced in a given 
annihilation event. The calculations were done for the case of $t_{\rm ann}=0$.

In the absence of FSI, $M_{\rm inv}$ should be equal to the invariant energy 
$\sqrt{s}$ of the annihilating $\bar p$N pair. Indeed, without FSI  and without 
mean field, as expected, we get a quite sharp peak at $2m_N$ only slightly
smeared out due to the Fermi motion of nucleons and momentum spread of
the initial antiproton DF (\ref{cohState}). The baryonic 
mean field leads to the shift of a peak position toward smaller $M_{\rm inv}$
and to some broadening of the distribution. The broadening is due to the spatial 
spread of the initial antiproton DF (\ref{cohState}), 
which results in different mean-field potentials acting on different annihilating 
$\bar p$N test particle pairs. Additionally, the FSI leads to a very strong 
broadening of the invariant mass spectrum due to the deceleration and 
absorption of the annihilation mesons. This is clearly seen in 
Fig.~\ref{fig:Minv_spectra_pbarO16} for the case when the RMF was switched
off (dotted line).
Nevertheless, the full calculation (solid line) shows quite strong softening
of the $M_{\rm inv}$ distribution due to the mean-field effects. 
Obviously, this effect is stronger for the case of $\xi=1$ as compared
with $\xi=0.3$. 

\begin{figure}

\includegraphics[bb = 64 75 590 730, scale = 0.9]{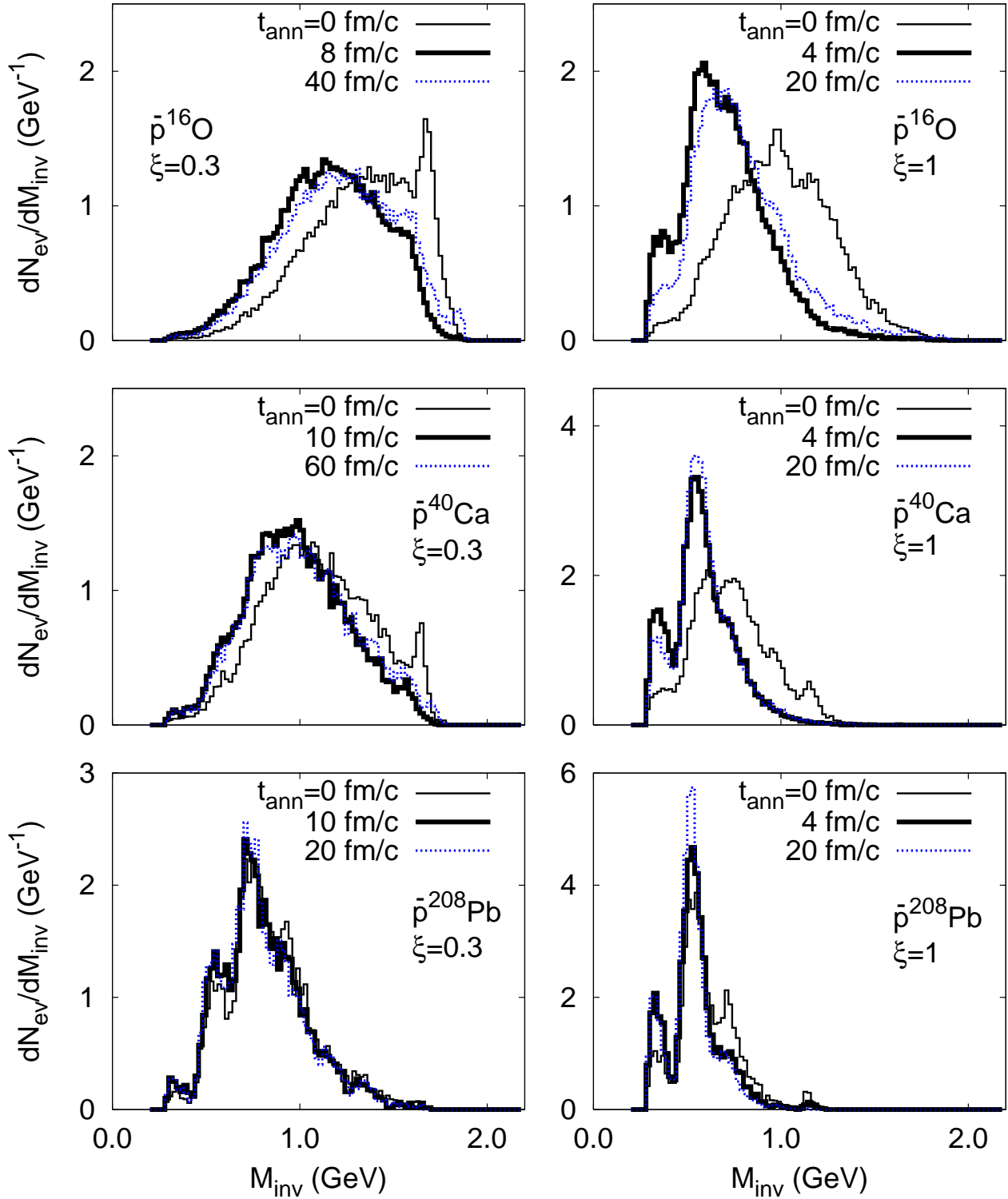}

\vspace*{0.5cm}

\caption{\label{fig:Minv_spectra} (color online) Annihilation event distributions on 
the total invariant mass of emitted mesons for various $\bar p$A systems and values of 
the parameter $\xi$. Different histograms correspond to different values of the annihilation time
$t_{\rm ann}$ indicated in the key.}

\end{figure}

Finally, in Fig.~\ref{fig:Minv_spectra}, we systematically study 
how the meson invariant mass spectra are affected by the nuclear 
compression effects. The results are shown for the different 
$\bar p$-nucleus systems. Due to the strong reduction of the nucleon
effective mass with the scalar density, the meson invariant mass spectra 
become softer when the annihilation happens in the compressed configurations, 
as compared  with the annihilation in the normal state at $t_{\rm ann}=0$. The 
effect is, again, more pronounced for the light systems $\bar p$$^{16}$O 
and $\bar p$$^{40}$Ca. In the $\bar p$$^{16}$O system the shift is almost
500 MeV even in the case of $\xi=0.3$.

We believe that results presented in 
Figs.~\ref{fig:Ekin_spectra},\ref{fig:pimul_pbarO16},\ref{fig:pimom_pbarO16} and
\ref{fig:Minv_spectra} constitute the set of observables sensitive to the compressional
effects in nuclei induced by an antiproton.

\section{ Summary and outlook }

We have performed dynamical modeling of possible compression 
effects in nuclei due to the presence of an antiproton. The 
semiclassical transport GiBUU model \cite{GiBUU} incorporating the relativistic
mean fields for the baryons and antibaryons has been employed in calculations. 
The model reproduces reasonably well the earlier static calculations of bound 
$\bar p$-nuclear systems \cite{Buer02,Mish05}.

In this work, we did not consider the stopping process of an incident antiproton 
in a target nucleus. This is a rather complicated problem due to the unknown in-medium 
cross sections of the $\bar p$-scattering and annihilation. This problem will be addressed 
in a forthcoming paper. Instead, we have assumed, that the antiproton has penetrated to the 
center of the nucleus, stopped there due to an inelastic collision, and then get captured to 
the lowest energy state. Such events should be very rare, with a probability of the order of 
$10^{-4}$ for the central collisions \cite{Mish05}. As proposed in Ref. \cite{Mish05}, formation 
of bound antiproton-nucleus states can be triggered by the emission of fast nucleons, pions or kaons.
We have shown, that during the time interval of $4\div10$ fm/c after creation of the initial 
state the central density of the target nucleus grows up to the values of $2\div3\rho_0$ 
depending on somewhat uncertain values of the antiproton coupling constants.
We expect that the life time of strongly bound antiprotons can be long enough 
to observe this cold compression effect. 

Detailed kinetic simulations of the post-annihilation evolution
of residual nuclei have been carried out at different assumptions on the annihilation time.
It is shown, that the $\bar p$-annihilation in compressed light systems, like $\bar p$$^{16}$O
and $\bar p$$^{40}$Ca, leads to the pronounced collective expansion of the residual
nucleus, which may result in the multifragment break-up. Another clear signature of
the nuclear compression is the hardening of the kinetic energy spectra of emitted
nucleons. On the other hand, the invariant mass distribution of produced mesons is shifted to
smaller invariant masses due to the in-medium reduction of the nucleon effective mass at high 
scalar density. Similar phenomena are expected also for the case of $\bar\Lambda$-nucleus bound 
states which can be produced via the $\bar p p \to \bar\Lambda \Lambda$ reaction on nuclei.

Another interesting possibility is that the compressed zone of the nucleus might
undergo a deconfinement phase transition. Then one can expect formation of a quark-antiquark
droplet with a non-zero baryon number and relatively low temperature \cite{Mish05}.  

Our main assumption in the present study was that the annihilation takes place 
in the central region of a nucleus. The experimental selection 
of the central annihilation events is a difficult problem. No clear trigger
condition for such events has been invented so far. One suggestion is that the
central annihilation events, in average, will be characterised by isotropic
emission of secondary particles and high fragment multiplicity \cite{Rafelski80,Cahay82}. 
However, further theoretical and experimental efforts are needed to develop a good trigger 
condition for the central annihilation. Despite of these difficulties, we propose to study
the above predictions in antiproton-nucleus reactions at the future Facility for 
Antiproton and Ion Research (FAIR) at GSI (Darmstadt).

\begin{acknowledgments}
The support by the Frankfurt Center for Scientific Computing is gratefully 
aknowledged. The authors are grateful to A.S. Botvina, Th.J. B\"urvenich, 
I.A. Pshenichnov, J. Ritman and H. St\"ocker for stimulating discussions. 
This work has been partially supported by the DFG Grant 436 RUS 113/711/0-2
(Germany) and the Grant NS-3004.2008.2 (Russia).
\end{acknowledgments}

\clearpage

\thispagestyle{empty}

\begin{table}[htb]
\caption{\label{tab:fragmenting_source} Fragmenting source parameters for
the different annihilating systems and values of the scaling factor $\xi$
of the antiproton coupling constant. $t_{\rm ann}$ denotes the annihilation
time moment. $t$ is the time moment when the source parameters have been determined.
$\rho_N$ is the central nucleon density.
$Z$, $E_{\rm kin}^{\rm coll}$ and $E_{\rm res}^\star$ are the charge
number, collective kinetic energy per nucleon and residual excitation
energy per nucleon, respectively (see Eqs. (\ref{NZ}),(\ref{E_kin^coll})
and (\ref{E_res^coll})).}

\vspace{1cm}

\begin{tabular}{|c|c|c|c|c|c|c|c|c|}
\hline
  System & $\xi$ & $t_{\rm ann}$ & $t$    & $\rho_N$    & Z & $E_{\rm kin}^{\rm coll}$ & $E_{\rm res}^\star$ \\ 
         &       & (fm/c)        & (fm/c) & (fm$^{-3}$) &   & (MeV/nucleon)            & (MeV/nucleon)\\
\hline
  ${\bar p}$$^{40}$Ca & 0.3 & 10 & 40 & 0.086 & 17 & 0.6 & 8.1 \\
  ${\bar p}$$^{40}$Ca & 1.0 & 4  & 40 & 0.071 & 16 & 0.3 & 6.3 \\
  ${\bar p}$$^{16}$O  & 0.3 & 8  & 35 & 0.057 & 6  & 2.0 & 9.9 \\
  ${\bar p}$$^{16}$O$^\ddag$ & 0.3 & 8  & 35 & 0.056 & 6  & 2.1 & 9.6 \\
  ${\bar p}$$^{16}$O  & 1.0 & 4  & 36 & 0.051 & 5  & 1.4 & 7.9 \\
  ${\bar p}$$^{16}$O$^\ddag$  & 1.0 & 4  & 36 & 0.051 & 5  & 1.5 & 7.8 \\
\hline
\end{tabular}
\end{table}

$^\ddag$ Including Coulomb interaction in propagation of the test particles.

\begin{table}[htb]
\caption{\label{tab:slopes} Slope temperatures $T_N$ (MeV) for the nucleon kinetic
energy spectra (see Fig.~\ref{fig:Ekin_spectra} and Eq. (\ref{fit_nuc})). Only the values 
of $T_N$ for $t_{\rm ann}=0$ (first number) and for the annihilation at the time of the maximum
compression (second number) are given. Statistical error is $\pm2$ MeV.}

\vspace{1cm}

\begin{tabular}{|c|c|c|c|}
\hline
   $\xi$   &  ${\bar p}$$^{16}$O &  ${\bar p}$$^{40}$Ca & ${\bar p}$$^{208}$Pb \\
\hline
   0.3     &  66, 81             &  67, 71              & 64, 59 \\
   1       &  52, 95             &  46, 79              & 45, 53 \\
\hline
\end{tabular}
\end{table}

\clearpage

\thispagestyle{empty}

\clearpage

\thispagestyle{empty}

\clearpage

\thispagestyle{empty}

\clearpage

\thispagestyle{empty}

\clearpage

\thispagestyle{empty}

\clearpage

\thispagestyle{empty}

\clearpage

\thispagestyle{empty}

\clearpage

\thispagestyle{empty}

\clearpage

\thispagestyle{empty}

\clearpage

\thispagestyle{empty}

\clearpage

\thispagestyle{empty}

\clearpage

\thispagestyle{empty}

\end{document}